\RequirePackage{amsmath}

\documentclass[epj]{svjour}

\usepackage{amssymb}
\usepackage{lipsum}
\usepackage[english]{babel}
\usepackage{graphicx}
\usepackage{booktabs}
\usepackage{multirow}
\usepackage{fancyref}
\usepackage[numbers,sort&compress]{natbib}
\usepackage{xcolor}
\usepackage{hyperref}
\usepackage[normalem]{ulem}
\hypersetup{
	colorlinks   = true,    
	urlcolor     = blue,    
	linkcolor    = blue,    
	citecolor    = blue     
}
\usepackage{caption}
\usepackage{subfig}

\newcommand {\bfp} {{\bf p}}
\newcommand {\bfq} {{\bf q}}
\newcommand {\bfr} {{\bf r}}

\renewcommand {\d} {{\rm d}}
\newcommand{\myin}{\mathrm{i}}
\newcommand{\myout}{\mathrm{o}}

\newcommand {\E} {\varepsilon}

\newcommand {\Om} {\Omega}

\begin{document}


\title{Transport of secondary electrons through coatings of ion-irradiated metallic nanoparticles}

\author{Kaspar Haume \inst{1,2}
	\and Pablo de Vera \inst{2,3}
	\and Alexey Verkhovtsev \inst{2,4,}
    \thanks{\email{verkhovtsev@mbnexplorer.com}; On leave from A.F. Ioffe Physical-Technical Institute, 194021 St. Petersburg, Russia}
	\and Eugene Surdutovich \inst{5}
	\and Nigel J. Mason \inst{1}
	\and Andrey V. Solov'yov \inst{2,}
	\thanks{On leave from A.F. Ioffe Physical-Technical Institute, 194021 St. Petersburg, Russia}
}                     
\institute{
	School of Physical Sciences, The Open University, MK7 6AA Milton Keynes, UK
	\and MBN Research Center, Altenh\"oferallee 3, 60438 Frankfurt am Main, Germany
	\and School of Mathematics and Physics, Queen's University Belfast, BT7 1NN Belfast, UK
	\and Instituto de F\'{\i}sica Fundamental, CSIC, Serrano 113-bis, 28006 Madrid, Spain
	\and Department of Physics, Oakland University, Rochester, 48309 Michigan, USA
}
\date{}

\abstract{
The transport of low-energy electrons through the coating of a radiosensitizing metallic nanoparticle under fast ion irradiation is analyzed theoretically and numerically.
As a case study, we consider a poly(ethylene glycol)-coated gold nanoparticle of diameter 1.6~nm excited by a carbon ion in the Bragg peak region in water as well as by more energetic carbon ions.
The diffusion equation for low-energy electrons emitted from a finite-size spherical source representing the surface of the metal core is solved to obtain the electron number density as a function of radial distance and time.
Information on the atomistic structure and composition of the coating is obtained from molecular dynamics simulations performed with the MBN Explorer software package.
Two mechanisms of low-energy electron production by the metallic core are considered: the relaxation of plasmon excitations and collective excitations of valence $d$ electrons in individual atoms of gold.
Diffusion coefficients and characteristic lifetimes of electrons propagating in gold, water, and poly(ethylene glycol) are obtained from relativistic partial wave analysis and the dielectric formalism, respectively.
On this basis, the number of electrons released through the organic coating into the surrounding aqueous medium and the number of hydroxyl radicals produced are evaluated.
The largest increase of the radical yield due to low-energy electrons is observed when the nanoparticle is excited by an ion with energy significantly exceeding that in the Bragg peak region. It is also shown that the water content of the coating, especially near the surface of the metal core, is crucial for the production of hydroxyl radicals.
}

\authorrunning{K. Haume et al.}
\titlerunning{Transport of secondary electrons through coating of ion-irradiated metallic nanoparticles}

\maketitle

\section{Introduction\label{sec:Introduction}}

Metal nanoparticles (NPs) made of gold, platinum as well as other metals like gadolinium or silver
have been examined as novel agents for more efficient treatment of tumors with ionizing radiation \cite{Kwatra2013,Bergs_2015_BiochimBiophysActa.1856.130, Yamada_2015_WIREs_Nanomed.7.428, Haume_2016_CancerNanotechnol.7.8}.
These NPs have attracted increasing interest because of their capacity to enhance the biological damage induced by
energetic photon and ion-beam irradiation.
Exposed to radiation such NPs can act as radiosensitizers \cite{Hainfeld2004, Polf2011, McMahon2011a, Porcel2010}, i.e.
they interact with radiation and produce a large number of secondary electrons, which may locally enhance damage of tumor cells relative to surrounding tissue.
The radiosensitizing potential of gold, platinum and gadolinium-containing NPs was demonstrated in experiments with plasmid DNA molecules \cite{Porcel2010, Porcel_2010_JPCS.373.012006, Xiao2011, Schlatholter2016} where an increased amount of single and double strand breaks was observed under photon, electron and ion irradiation.
Several \textit{in vitro} experiments with living cells irradiated with photons and ions also provided evidence of enhancement of radiation-induced effects in the presence of metal NPs \cite{Porcel2014a, Li_2016_Nanotechnology.27.455101}.

It is now accepted that the main pathway of biological damage induced by ionizing radiation is mediated by secondary electrons and free radicals \cite{Nano-IBCT2017,Surdutovich2014, Sanche_2005_EPJD.35.367, Alizadeh_2015_AnnuRevChemPhys.66.379} and the
radiosensitization by metallic NPs is therefore commonly related to enhanced emission of secondary electrons~\cite{Xiao2011} which activate hydrolysis of the surrounding water medium and facilitate radical production~\cite{Sicard-Roselli2014}. In particular, low-energy electrons (LEEs) with energies up to several tens of eV are recognized as essential agents of biodamage by ion beams \cite{Nano-IBCT2017,Surdutovich2014, Sanche_2005_EPJD.35.367} and, as such, the focus of this work is centered on them.

Nanoparticles in biomedical applications are usually synthesized with an organic coating to improve stability under physiological
conditions, reduce toxicity and target specific biological sites.
These biological aspects are a subject of intense research with numerous experiments being performed both \textit{in vitro} and \textit{in vivo} \cite{Yamada_2015_WIREs_Nanomed.7.428,Haume_2016_CancerNanotechnol.7.8, Rosa_2017_CancerNanotechnol.8.2}.
Apart from the choice of a coating, a variation in electronic and magnetic properties of metallic NPs may be important
for biomedical applications \cite{McNamara_2017_AdvPhysX.2.54}. For instance, magnetic NPs can be used for magnetic resonance imaging applications or can be delivered to the tumor region being guided by a magnetic field \cite{Klein_2014_JPCB.118.6159}.

In general, a vast number of parameters can be varied to optimize the radiosensitizing properties of NPs (e.g., the size, shape and composition of a metal core; thickness, composition and density of the coating) which makes it a formidable task to systematically explore each of them experimentally.
As a consequence computational modeling is often utilized to evaluate the potential of a given system and to shed light on the molecular-level mechanisms underlying radiosensitization by NPs before experiments are carried out.

Transport of electrons and free radicals in the vicinity of metal NPs
is frequently explored by means of Monte Carlo simulations \cite{Prezado_2015_MedPhys, Lin2014, McMahon_2016_Nanoscale, Tran_2016_NIMB}. According to the outcomes of these simulations
metal NPs irradiated with energetic photon beams emit a large number of secondary electrons via the Auger mechanism. However, Auger cascades are less relevant for ion beams due to much lower probabilities of ions to ionize inner shells of atoms in the target \cite{Heredia-Avalos2007d}. In this work we consider irradiation with ions in the Bragg peak region in water (energy of 0.3~MeV/u) as well as outside the Bragg peak region (energies from 1 to 10~MeV/u). For most of these energies the electron emission channel associated with ionization of inner shells in the atoms of metals (e.g., gold) and subsequent emission of Auger electrons is suppressed, both due to the small cross sections for ionization of inner shells and to the limits imposed by the kinematic limit, with the maximum possible energy transfer often being smaller than the ionization thresholds of inner shells. Therefore, in this work we focus our analysis on other important channels of electron emission from outer electronic shells, which are associated with collective electronic excitations in a metallic target.

The majority of Monte Carlo simulations conducted so far have considered ``naked'' metal NPs (without  coating).
However earlier experimental studies have indicated that a coating may affect the biodamage induced by NPs exposed to radiation \cite{Gilles2014,Xiao2011}.
Therefore, the impact of a coating layer on the production of secondary electrons and free radicals is an important scientific question which should be addressed in simulations before the potential of NPs as radiosensitizing agents can be accurately estimated.

The aforementioned problem is complex because of the actual system sizes, molecular interactions, radiation dynamics and their links to biological effects.
A realistic approach to tackle such a complex problem should involve the theoretical descriptions of the key phenomena and elaborate their major interlinks within a unifying framework. Such an approach has been developed during the past decade for the description of the molecular-level mechanisms of biological damage induced by ion-beam irradiation.
The most recent review of these studies is given in reference \cite{Nano-IBCT2017}.
These studies emphasized the cross-disciplinary nature of the problem and led to the formulation of the multiscale approach to the physics of radiation damage with ions (see Refs.~\cite{Nano-IBCT2017, Surdutovich2014} and references therein).
One of the important achievements of this approach concerns the possibility to predict survival probabilities for living cells irradiated with ions \cite{Verkhovtsev2016} through a quantitative description of the key physical, chemical, and biological phenomena which occur over different time, space and energy scales.
As such, the multiscale approach explicitly includes ionization of the medium by ion projectiles, formation and transport of secondary particles, chemical interactions, thermomechanical pathways of biodamage, and heuristic biological criteria for cell survival \cite{Surdutovich2014}.

In this paper, we present a novel theoretical and computational approach to analyze electron emission from coated metallic NPs irradiated with ions.
The methodology developed is general and can be applied for NP of different size and coating thickness as well as for any combination of metallic core and organic coating. As an illustrative case study we consider a gold NP of diameter 1.6~nm coated with poly(ethyle\-ne glycol) (denoted Au@PEG) -- one of the most commonly used coating materials.
We consider a coating composed of PEG molecules of five sub-units.
Structure of such NP was explored recently by means of molecular dynamics (MD) simulations \cite{Haume2016}.
We evaluate the number of electrons that penetrate through the coating after their emission from the metallic core induced by a passing C$^{6+}$ ion of 0.3~MeV/u energy. Carbon ions are currently used in ion-beam cancer therapy, one of the most promising radiotherapies currently available \cite{Nano-IBCT2017}.
The energy of 0.3~MeV/u corresponds to the Bragg peak region of the ion's trajectory in water, which is the region of maximal energy deposition. This analysis is also extended towards higher ion energies outside the Bragg peak region, namely from 1 to 10 MeV/u.

Two mechanisms of LEE production by the metallic core are considered; the first is associated with a plasmon excitation and the second with a collective excitation of valence $d$ electrons in individual atoms of gold.
As demonstrated in Refs.~\cite{Verkhovtsev2015a, Verkhovtsev2014a}, these two mechanisms play a significant role in the enhanced LEE emission from ``naked'' metallic NPs irradiated with ions.
The yield of electrons with the energy of about $1 - 10$~eV is strongly increased due to the decay of plasmon-type collective excitations that involve valence electrons delocalized over the whole NP.
In particular a significant contribution to the LEE yield due to plasmon excitations comes from the surface plasmon; for a few-eV electrons emitted from small (of 1~nm diameter) NPs, its contribution to the ionization cross section is about an order of magnitude higher than that of the volume plasmon \cite{Verkhovtsev2015a}.
For higher electron energies (of a few tens of eV) the main contribution to the electron yield arises from the atomic giant resonance associated with the collective excitation of valence $d$-electrons in individual atoms of a NP.

\begin{figure}[htb!]
    \begin{centering}
    \includegraphics[width=0.36\textwidth]{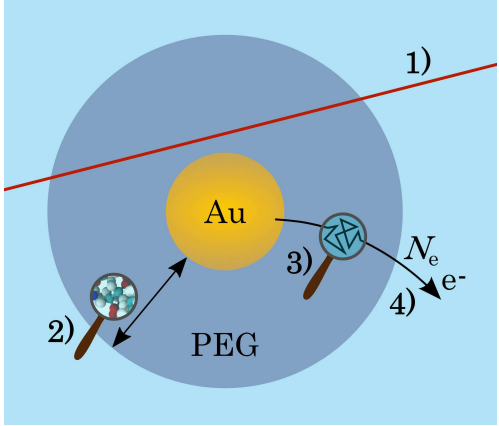}
    \par\end{centering}
    \caption{We consider an Au@PEG NP in water excited by a passing ion and apply the following workflow. As a result of ion irradiation (1), $N_\mathrm{e}$ electrons are emitted from the metal core due to plasmons and collective electron excitations in individual atoms of gold. After simulating the structure of the coated NP by means of molecular dynamics (2), the interaction mechanisms between electrons and coating (diffusion coefficients and average lifetimes) are quantified by calculating elastic and inelastic scattering mean free paths (3). These numbers are used to obtain the number density of electrons passing through the coating (4) by solving the diffusion equation.}
    \label{fig:outline}
\end{figure}

The dipole and higher multipole excitation modes can contribute to the LEE production by the plasmon-type excitations \cite{Solovyov_review_2005_IntJModPhys.19.4143, Gerchikov_1997_JPB.30.4133}. The dipole mode is excited when the characteristic impact parameter
significantly exceeds the radius of the core, $v_{\rm ion}/\Delta\E \gg R$\footnote{Here and in Section~\ref{sec:pra} we use mainly the atomic system of units, $m_e = |e| = \hbar = 1$, unless otherwise indicated.} \cite{Solovyov_review_2005_IntJModPhys.19.4143, Gerchikov_1997_JPB.30.4133}. Here $v_{\rm ion}$ is the projectile velocity and $\Delta\E$ is a characteristic energy transferred during the collision ($\Delta\E \approx 5$~eV for a plasmon-type excitation in a gold NP). For the ion's energy of 0.3~MeV/u this characteristic impact parameter is about 0.9~nm, and the radius of the metallic core which we consider as a case study is $R = 0.8$~nm.
In this case, $v_{\rm ion}/\Delta\E \gtrsim R$ and the dipole, quadrupole and octupole plasmon excitation modes contribute to the LEE production.

The dipole mode of collective $5d$-electron excitations in individual atoms of a gold NP will be excited in the atoms that are confined within a cylinder with radius $r \lesssim v_{\rm ion}/\Delta\E$ from the ion's path, where $\Delta\E \approx 25$~eV \cite{Verkhovtsev2015a}. An estimate for a 0.3~MeV/u-ion gives $r \le 0.25$~nm which means that the excitations of $5d$ electrons are formed in the atoms positioned no further than 0.25~nm from the ion's path. This distance is smaller than the lattice constant of bulk gold, suggesting that only a few atoms positioned in close proximity to the ion's track will be excited via this mechanism. However, as shown in this paper, this mechanism contributes significantly to LEE emission when the energy of a projectile ion is considerably larger than that in the Bragg peak region in water and corresponds to the entrance segment of ion's trajectory.

The general workflow applied in this paper is schematically shown in Fig.~\ref{fig:outline}. A diffusion equation-based approach \cite{Surdutovich2015} is used to describe the transport of electrons emitted from the NP core through the coating. The number of electrons emitted from the metal core due to collective electron excitations as a result of ion irradiation is evaluated by means of the formalism described in Refs.~\cite{Verkhovtsev2015a, Verkhovtsev2014a}.
Information on the NP core and the coating structure is obtained from MD simulations \cite{Haume2016}. The inelastic scattering cross sections between electrons and the coating/water compound material are evaluated by means of the dielectric formalism \cite{Abril1998,Heredia-Avalos2005,Abril2013,deVera2013} while elastic scattering is treated within relativistic partial-wave analysis \cite{Salvat2003,Salvat2005}.

We demonstrate that the majority of emitted electrons undergo an inelastic collision before escaping the coating.
From MD simulations of the water content in the coating layer, the probability of an electron targeting a water molecule is extracted and the OH radical yield from inelastic collisions of the electrons with water molecules is quantified.
It is shown that the number of hydroxyl radicals produced in the vicinity of Au@PEG NP is about $2-7$ times smaller (depending on the projectile ion energy and the coating mass density) than that in the vicinity of a ``naked'' gold NP. This observation is in agreement with earlier experimental results \cite{Gilles2014}, suggesting that radiosensitization efficiency of metallic NPs is affected by the thickness and water content of the coating.
Lastly, we demonstrate that the production of OH radicals is enhanced by an order of magnitude when the NP is irradiated by $5-10$~MeV/u C$^{6+}$ ions as compared to the case of 0.3~MeV/u irradiation.

The paper is outlined as follows. In Section~\ref{sec:Methodology} we present the general metho\-dology of this work.
This is followed by a discussion of the results in Section \ref{sec:results}. Finally, the findings of this study are summarized in Section \ref{sec:conclusion}.
In Appendix \ref{Appendix_Diff_eq}, we present a complete derivation of the diffusion equation accounting for the geometry and the initial conditions relevant for this work.
In Appendix \ref{app:cross_sections}, the methodology for calculating interaction cross sections and mean free paths is described in greater detail.

\section{Methodology \label{sec:Methodology} }
\subsection{Solution of the diffusion equation \label{sec:diffusion} }

In Ref.~\cite{Surdutovich2015} the transport of LEEs and reactive species (free radicals and solvated electrons) brought about by ions traversing liquid water in the vicinity of the Bragg peak was studied by means of the diffusion model.
The production of energetic $\delta$-electrons is suppressed in the Bragg peak region \cite{deVera_2017_EPJD} and the diffusion approach is therefore suitable for describing the transport of the product electrons which will be of low energy.
The reliability of the diffusion equation-based approach for modeling the LEE transport in liquid water was examined through the comparison with track-structure Monte Carlo simulations and a good agreement between the two approaches was reported (see Refs.~\cite{Nano-IBCT2017, Surdutovich2014} and references therein).
The use of the diffusion model relies on the assumption that the angular dependence of elastic and inelastic scattering cross sections, which govern interactions of LEEs with molecules of the medium, is rather weak. Therefore, these processes can be considered as isotropic to a first approximation.

In this work we consider a metal core of a Au@PEG NP irradiated by an ion as a source of electron emission. The diffusion model is utilized to study the transport of LEEs emitted from a spherically symmetric core of radius $R$  through the coating. Let us consider that $N_\mathrm{e}$ electrons are emitted from the surface at the time instance $t$ -- we refer to these electrons as first-generation electrons. The passage of a $0.3 - 10$~MeV/u projectile along the NP can be considered as a fast process (the ion velocity $v_{\rm ion} > 1$~a.u.). Therefore, one can refer to this instance as a moment when the secondary electrons are emitted from the surface of the metallic core and their post-collision, relatively slow diffusion process begins. Thus, this instance  corresponds to $t = 0$ in the considered diffusion process.
We apply the developed model for a particular case study of a NP with the core diameter of 1.6~nm taken from MD simulations (see Sect.~\ref{sec:MD}). This size is comparable to the mean free path of electrons emitted due to collective $5d$-electron excitations.
It is therefore meaningful to assume, in the first approximation, that all electrons are ejected from the surface of the metal core. A more rigorous analysis of electron emission from the bulk region of the core can also be performed but this task goes beyond the scope of this work.

It is assumed that the emitted electrons can propagate in any direction i.e. through the coating as well as inside the core.
The three-dimensional diffusion of electrons is then described by the following equation:
\begin{equation}
\frac{\partial n_1(\bfr,t)}{\partial t} = D_1 \nabla^2 n_1(\bfr,t) - \frac{n_1(\bfr,t)}{\tau_1} \ ,
\label{Diff_eq_01}
\end{equation}
where $n_1(\bfr, t)$ is the number density of first-generation electrons at point $\bfr$ and time instance $t$.
The diffusion coefficient $D_1$ is related to the electrons' velocity $v_1$ and their elastic scattering mean free path (MFP) in the medium $\lambda_{1, \rm el}$ as $D_1 = v_1 \lambda_{1, \rm el}/6$. The second term on the r.h.s. of Eq.~\eqref{Diff_eq_01} accounts for attenuation of electrons experiencing inelastic collisions.
The characteristic lifetime of the first-generation electrons is given by the constant $\tau_1$ which is related to the inelastic MFP $\lambda_{1, \rm inel}$ as $\tau_1 = \lambda_{1, \rm inel}/ v_1$.
Taking into account the spherical symmetry of the problem and
the different values of $D_1$ and $\tau_1$ in the ``inner'' region (the metallic core, $0<r<R$) and in the ``outer'' region (coating and the surrounding medium, $r \geq R$), Eq.~\eqref{Diff_eq_01} transforms into
\begin{multline}
\frac{\partial n_1(r,t)}{\partial t} =  \\
\left\{
\begin{array}{l l}
D_{1, \rm i} \,
\displaystyle{
\frac{1}{r^2}
\frac{\partial}{\partial r} \left( r^2 \frac{\partial n_1(r,t)}{\partial r} \right) -
\frac{n_1(r,t)}{\tau_{1, \rm i}} } \ , & 0 < r < R
\vspace{0.1cm} \\
\displaystyle{
D_{1, \rm o} \, \frac{1}{r^2}
\frac{\partial}{\partial r} \left( r^2 \frac{\partial n_1(r,t)}{\partial r} \right) -
\frac{n_1(r,t)}{\tau_{1, \rm o}} } \ , & r \geq R  \ .
\end{array}
\right.
\label{Diff_eq_02}
\end{multline}

The solution of Eq.~\eqref{Diff_eq_02} should satisfy the following initial and boundary conditions:
(i) all the electrons are emitted simultaneously at the initial time ($t = 0$) and from the surface with the radius~$R$,
\begin{equation}
n_1(r \ne R, 0) = 0, \qquad \int_0^{\infty} n_1(r, 0) \, 4\pi r^2 {\rm d}r = N_{\rm e} \ ,
\end{equation}
and (ii) the number density of electrons is equal to zero at large distances from the source,
\begin{equation}
n_1(r\to \infty, t) \to 0 \ ,
\end{equation}
see Appendix \ref{Appendix_Diff_eq} for more details.

An Au@PEG NP embedded in water is an inhomogeneous system where electron transport through the metal core and through the coating medium is characterized by different values of $D$ and $\tau$.
In the general case the analytic solution of the diffusion equation given by Eq.~\eqref{Diff_eq_02} is not possible.
However, it is feasible for a homogeneous problem where the two aforementioned regions are characterized by the same values of diffusion coefficient $D_\myin = D_\myout \equiv D$ and electron lifetime $\tau_\myin = \tau_\myout \equiv \tau$.
The analytic solution reads as
\begin{multline}
n_1(r,t) =  \frac{N_\mathrm{e}}{8\pi R r \sqrt{\pi D_1 t}}
\exp \left( - \frac{t}{\tau_1} \right)  \\
\times
\left[ \exp \left( - \frac{(r - R)^2}{4D_1 t} \right)
- \exp \left( - \frac{(r + R)^2}{4D_1 t}  \right)
\right] \ .
\label{number_dens_1st_gen}
\end{multline}
Details of the derivation procedure are described in Appendix~\ref{Appendix_Diff_eq_analytic}.

The general solution of Eq.~\eqref{Diff_eq_02}, accounting for the different values of $D_1$ and $\tau_1$ in the inner and outer regions, was found numerically.
More details of this numerical procedure are presented in Appendix~\ref{Appendix_Diff_eq_2media}.

In a similar way one can account for the second generation of electrons~\cite{Surdutovich2015} which are produced as a result of ionization of the medium by electrons emitted from the NP core (the first generation). If the energy of the latter exceeds the ionization potential of a PEG or a water molecule, ionization events may take place resulting in the decay of the first-generation electrons and the production of two second-generation electrons.

The general diffusion equation for the second generation of electrons is given by
\begin{multline}
\frac{\partial n_2(r,t)}{\partial t} = D_2 \frac{1}{r^2}
\frac{\partial}{\partial r} \left( r^2 \frac{\partial n_2(r,t)}{\partial r} \right)
- \frac{n_2(r,t)}{\tau_2} \\
+ 2 \frac{n_{1, E > I_p}(r,t)}{\tau_1} \ .
\label{Diff_eq_2nd_gen}
\end{multline}
where the subscripts ``1'' and ``2'' refer to electrons of the first and second generation, respectively.
The positive term $2 n_{1, E > I_p}(r,t)/\tau_1$ is a consequence of the fact that each electron from the first generation whose energy $E$ exceeds ionization potential $I_p$ of a water/PEG molecule and which undergoes an inelastic collision at a rate given by the average lifetime $\tau_1$ is transformed into two second-generation electrons.

The number density $n_2(r,t)$ can be calculated using the Green's function formalism:
\begin{align}
n_2(r, t) = 2 \int G_2 (\bfr - \bfr^{\prime}, t - t^{\prime}) \,
\frac{n_{1, E > I_p}(\bfr^{\prime},t)}{\tau_1} \, \d\bfr^{\prime} \d t
\end{align}
where $G_2(\bfr - \bfr^{\prime}, t - t^{\prime})$ is the three-dimensional Green's function which reads as
\begin{multline}
G_2 (\bfr - \bfr^{\prime}, t - t^{\prime}) =
\left( \frac{1}{4 \pi D_2 (t - t^{\prime}) } \right)^{3/2} \\
\times \exp \left( - \frac{ (\bfr - \bfr^{\prime})^2 }{ 4 D_2 (t - t^{\prime})} - \frac{t - t^{\prime}}{\tau_2} \right) \ 
\end{multline}

After rewriting $(\bfr - \bfr^{\prime})^2 = r^2 + r^{\prime 2} - 2rr^{\prime} \cos{\theta}$ due to the spherical symmetry of the problem and evaluating analytically the angular components of the integral, the number density of electrons created outside the metal core, $n_{2, \rm o}(r,t)$, is written as
\begin{align}
n_{2, \rm o}(r, t) &=
\frac{4}{\tau_{1, \rm o} r} \int_{R}^{\infty} \d r^{\prime} \int_0^t \d t^{\prime}
\left( \frac{1}{4 \pi D_{2, \rm o} (t - t^{\prime}) } \right)^{1/2}  \nonumber \\
&\times \exp \left( - \frac{ r^2 + r^{\prime 2} }{ 4 D_{2, \rm o} (t - t^{\prime}) }  - \frac{t - t^{\prime}}{\tau_{2, \rm o}} \right)  \nonumber \\
&\times n_{1,{\rm o},E > I_p}(r^{\prime}, t^{\prime}) \,
r^{\prime} \, \sinh{ \left( \frac{2 r r^{\prime}}{4 D_{2, \rm o} (t - t^{\prime}) } \right) }  \ ,
\label{eq:n2full}
\end{align}
where the integrals over $\d r^{\prime}$ and $\d t^{\prime}$ should be done numerically.

Using this methodology we neglect the contribution of the second-generation electrons that have been formed within the metal core and propagated to the coating region from the inside.
To accurately capture this non-local effect, one must derive a Green's function which takes into account the propagation of electrons in both the inside and the outside regions simultaneously.
This is a rather complex mathematical task that goes beyond the scope of the work presented here.
The accuracy of the overall calculation of the total number density of emitted electrons and the production of radicals should not be impaired by this approximation.

Knowing the number densities $n_1(r,t)$ and $n_2(r,t)$, one obtains the flux of electrons $\Phi(r,t)$, that is the number of particles propagating through a unit area per unit time. It can be found from Fick's first law of diffusion, $\Phi(r,t) = -D \, \partial{ n(r,t) }/\partial{r}$.
The fluence of electrons is calculated as the integral of the flux over the entire time after the emission.
Integrating the fluence over the surface of a sphere of radius $r$, the number of electrons passing through the coating at a given distance $r$ is given by:
\begin{align}
F(r) &= \int_0^{\infty} 4 \pi r^2 \, \Phi(r,t) \, \d t \ .
\label{eq:1stgen-fluence}
\end{align}

\subsection{Molecular dynamics of NP and PEG structure\label{sec:MD}}

The main ingredient for calculating the elastic and inelastic scattering MFPs, which determine $D$ and $\tau$, is the density distribution of atoms in the coating. This was calculated with molecular dynamics (MD) simulations using the MBN Explorer software package \cite{Solovyov2012, MBNExplorer_book, MBNExplorer_Tutorials_book}. The detailed procedure is described in Ref.~\cite{Haume2016}; a brief summary is given below.

\begin{figure*}[t]
	\begin{centering}
	\includegraphics[width=0.90\textwidth]{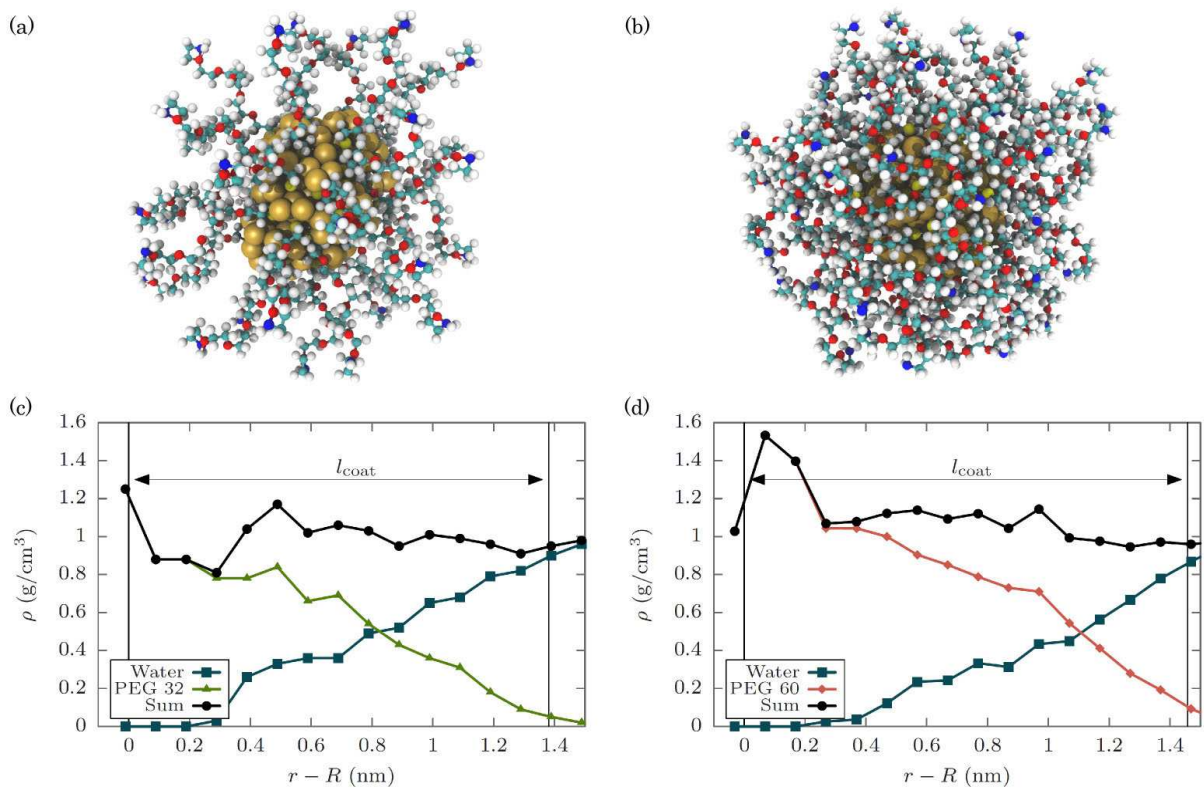}
	\par\end{centering}
    \caption{Illustrations of the 1.6~nm AuNP coated with (a) 32 and (b) 60 PEG molecules, obtained from MD simulations. The lower panels show the corresponding mass density $\rho$ of the combined elements of the PEG coating, the water content of the coating and their sum versus distance from the boundary of the NP core for (c) 32 and (d) 60 PEG molecules attached. The extent of the coating (vertical lines) and the resulting thickness $l_\mathrm{coat}$ are indicated as defined in the text.}
    \label{fig:densityDistr}
\end{figure*}

The metal core of the NP comprised 135 gold atoms, corresponding to a diameter of approximately 1.6~nm.
The core was coated with either 32 or 60 PEG molecules (5 ethylene glycol sub-units) altered such that they had a thiol group (used for bonding to the gold surface) on one end and an amino group on the other end.
The resulting structures of Au@PEG$_{32}$ and Au@PEG$_{60}$ NPs are shown in the upper panel of Fig.~\ref{fig:densityDistr}.
Each system was solvated explicitly in water at atmospheric pressure and equilibrated at 310~K.

The thickness of the coating was defined as $l_\mathrm{coat} = r_{97\%} - \left< r_\mathrm{S} \right>$, where $\left< r_\mathrm{S} \right>$ is the average distance of the sulfur atoms to the center of mass (CM) of the system and $r_{97\%}$ the distance from the CM inside which 97\% of the PEG atoms could be found. Both $\left< r_\mathrm{S} \right>$ and $r_{97\%}$ were taken after averaging over a number of frames of the MD simulation (see Ref. \cite{Haume2016} for details). The coating thickness $l_\mathrm{coat}$ for $N_\mathrm{PEG} = 32$ and 60 ($N_\mathrm{PEG}$ is the number of PEG molecules attached to the metal core) was approximately 1.4~nm in both cases.

The density distribution of the coating was calculated by counting the number of atoms of each element in concentric shells of thickness 1~\AA~around the CM and finding the corresponding mass density. The density distributions for the coatings composed of 32 and 60 PEG molecules are shown in the lower panel of Fig.~\ref{fig:densityDistr}. For both coatings, the total density is around 1.0~g/cm$^3$ except for fluctuations close to the NP boundary. These arise due to the fact that the NP core after annealing deviates from the perfect spherical symmetry, meaning that not all sulfur atoms are captured in the same shell when calculating the density. As the PEG density decreases, the water content increases correspondingly. This difference in number of attached PEG molecules then results in a difference in the degree of water penetration into the PEG coating. In the PEG 60 coating, the amount of water is thus significantly less than in the PEG 32 coating.

From the coating thickness and the density distribution analysis the average chemical composition, average density $\left< \rho \right>$, and mean atomic number $\left< Z_\mathrm{t} \right>$ of the coating were determined, see Table~\ref{tab:pegCoatings}. These values were used to calculate the MFPs as described in Section~\ref{sec:cross_sections}.
The data collected in Table~\ref{tab:pegCoatings} reveals that the main difference between the two coating media, relevant to the calculation of MFPs, is the resulting mean density which is about 10\% higher for $N_\mathrm{PEG} = 60$ due to the slightly higher density of PEG compared to water.

\begin{table}
	\centering
	\caption{The average chemical formula, mean density $\left< \rho \right>$, and mean atomic number of the coating $\left< Z_\mathrm{t} \right>$ versus number of attached PEG molecules $N_\mathrm{PEG}$ for the two coatings considered.}
	\label{tab:pegCoatings}
	\begin{tabular}{llll}
\hline		
			\multicolumn{1}{c}	{$N_\mathrm{PEG}$}	
		& 	\multicolumn{1}{c}	{Chemical formula }
		& 	\multicolumn{1}{c}	{$\left< \rho \right>$ (g/cm$^3$) }	
        & 	\multicolumn{1}{c}	{$\left< Z_{\mathrm{t}}\right>$ }  \\
\hline
32	& C$_{18.2}$H$_{146.4}$O$_{61.2}$N$_{1.4}$S$_{1.0}$	& 0.99	 & 3.38	\vspace{0.2cm} \\	
		  	
60	& C$_{16.9}$H$_{107.3}$O$_{42.4}$N$_{1.4}$S$_{1.0}$	& 1.08	 & 3.39 \\	
\hline
	\end{tabular}
\end{table}

\subsection{Interaction mean free paths of electrons \label{sec:cross_sections}}

As discussed above the diffusion coefficient of electrons, $D$, is related to the MFP of elastic scattering, $\lambda_\mathrm{el}$, while the average electron lifetime $\tau$ is determined by the inelastic MFP $\lambda_\mathrm{inel}$.
Therefore both elastic and inelastic scattering cross sections (CSs) for the electron collision with gold, PEG and liquid water have to be known to solve the diffusion equation~\eqref{Diff_eq_02}. The calculation of a reliable set of CSs for a wide energy range, including very low-energy electrons, is a complex task \cite{Emfietzoglou2005,Incerti2010,Blanco2013,Kyriakou2015} which should include advanced \textit{ab initio} approaches and is beyond the scope of this work.
Instead we aim here to obtain reasonable estimates for CSs and MFPs for the three materials and to emphasize the relative differences between them.

\subsubsection{Inelastic scattering mean free paths \label{sec:inelastic}}

In this work we consider electronic excitations and ionizations as the main contributor to energy loss of LEEs and complement it with an estimate for the CSs of vibrational excitation of the water and coating molecules, which is important for electrons of low energy \cite{Dingfelder2008c,Itikawa2005}.

Electronic interaction CSs for condensed phase materials are calculated within the dielectric formalism \cite{Abril1998,Heredia-Avalos2005,Lindhard1954,Ritchie1977,Abril2013}. This approach is based on the first Born approximation from the quantum point of view, or on the linear response of the electron density to the external electric field (i.e., polarization) from the classical point of view. Being a first order approximation it is applicable for fast electrons having energies larger than about 300~eV. At lower energies, exchange corrections due to the non-distinguishability of the primary and the emitted electrons are accounted for to increase the accuracy of the method \cite{Fernandez-Varea1993,deVera2011,Garcia-Molina2016}.

The main ingredient of the dielectric formalism is the energy-loss function (ELF) of the material, ${\rm Im}[-1/\epsilon(\Delta\E,q)]$, where $\epsilon(\Delta\E,q)$ is the complex dielectric function with $\Delta\E$ and $q$ being the energy and momentum transferred in the electronic excitation respectively. It represents the electronic excitation spectrum of the condensed phase material.
The ELF is usually obtained from experimental data in the optical limit ($q=0$) and then suitable dispersion algorithms are used to extrapolate the ELF over the whole energy and momentum plane \cite{Garcia-Molina2012b}.
Once the complete ELF is known the dielectric formalism allows the energy loss quantities of charged particles in condensed matter to be calculated \cite{Abril2013} including the electron production CSs \cite{deVera2013,deVera2015}.

It is difficult to find experimental information on the ELF of a specific coating considering the wide variety of possible compositions and the fact that the coating is partially mixed with water, which changes its density and overall composition.
Fortunately, the optical ELF of all organic materials is quite similar with an intense excitation peak around 20~eV. This has allowed the optical ELF of organic materials to be parameterized \cite{Tan2004} and predicted by just knowing the atomic composition and average density of the material.

\begin{figure}[t]
	\begin{centering}
	\includegraphics[width=0.45\textwidth]{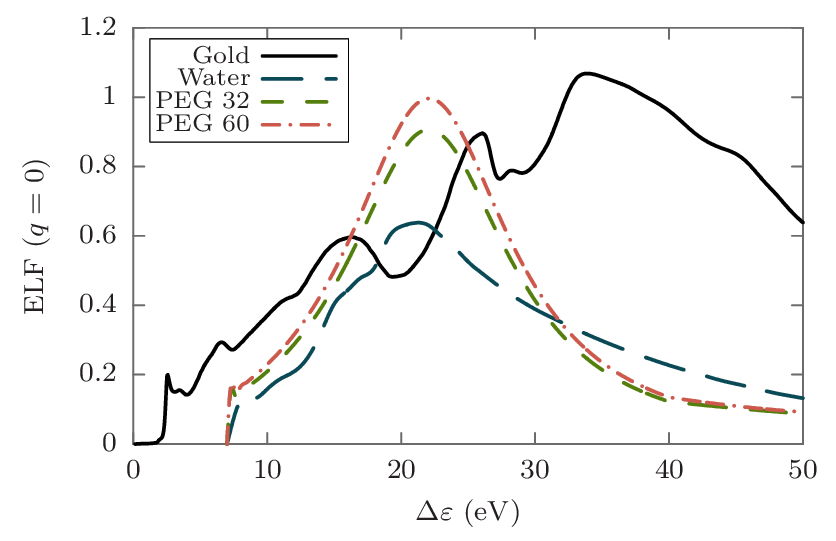}
	\par\end{centering}
    \caption{Energy-loss function of gold, liquid water, and two PEG coatings of different density (composed of 32 and 60 molecules) in the optical limit ($q = 0$). The ELF of liquid water is obtained from inelastic X-ray scattering data \cite{Hayashi2000} and for gold from its optical properties \cite{Palik1999}.}
    \label{fig:ELFs}
\end{figure}

\begin{figure}[t]
    \begin{centering}	
    \includegraphics[width=0.45\textwidth]{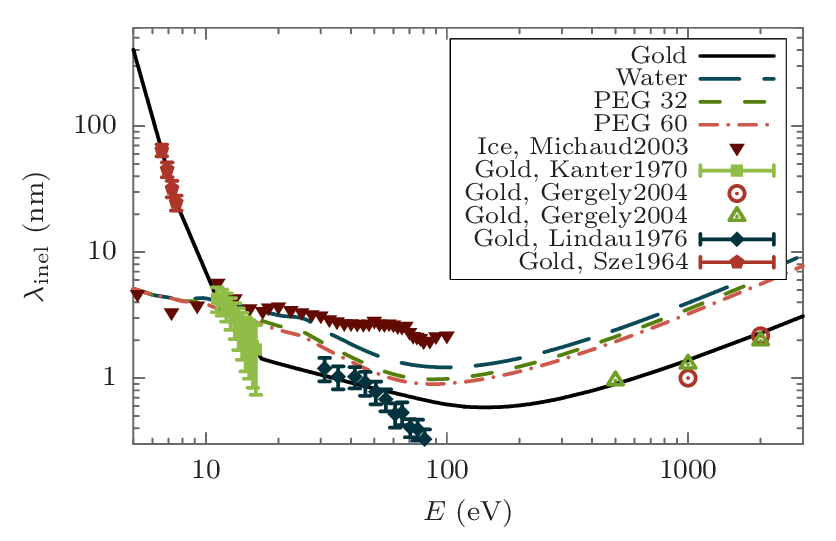}
    \par\end{centering}
    \caption{Inelastic scattering mean free path $\lambda_\mathrm{inel}$ of electrons in gold, liquid water and two PEG coatings. Lines represent the data compiled from calculations within the dielectric formalism (for high energies) and experimental data (for low energies) as explained in the text. Symbols are experimental data for ice \cite{Michaud2003} and gold \cite{Sze1964,Lindau1976,Kanter1970, Gergely2004}. }
    \label{fig:inelMFP}
\end{figure}

We have used this parametric approach to predict the optical ELF of the two PEG coatings, whose density and composition were obtained from the MD simulations (see Table~\ref{tab:pegCoatings}). These ELFs together with the data for liquid water \cite{Hayashi2000} and gold \cite{Palik1999} are shown in Fig.~\ref{fig:ELFs}.
The similarity between the PEG 32 and PEG 60 coatings is apparent with the slight difference being a result of a difference in density of about 10\%.
Note also that the ELFs of the coatings are qualitatively similar to that of liquid water, while the ELF of gold has a distinct behavior. The optical ELFs of these materials were extended for finite values of $q$, following Garcia-Molina, Abril \textit{et al.}, by means of the MELF-GOS (Mermin-Energy-Loss Function -- Generalized Oscillator Strengths) method \cite{Abril1998,Heredia-Avalos2005,Garcia-Molina2009a,Garcia-Molina2012b}. These results were used to obtain the inelastic MFP for electrons in the three materials within the dielectric formalism \cite{Abril2013,Garcia-Molina2012b,deVera2011} and using the Mott exchange factor \cite{ICRU55} as discussed in more detail in Appendix~\ref{app:inel_cross_sections}.
Similar calculations were reported previously in Refs.~\cite{Denton_2008_SurfInterfaceAnal.40.1481, Garcia-Molina2016}.

As discussed above the dielectric formalism loses accuracy for very low energy electrons. Furthermore, for water and organic molecules, other inelastic channels in addition to electronic interactions become increasingly important at low energies. Among them vibrational excitations represent one of the main inelastic channels. Due to these limitations for energies below $\sim 20$~eV we rely on experimental data for the inelastic MFP.

For gold our calculations agree rather well with experimental data down to $\sim 16$~eV (see Appendix~\ref{app:inel_cross_sections}). Below this energy we extended the calculated data with an interpolation of the very low energy experimental data \cite{Kanter1970,Sze1964}. The solid line in Fig.~\ref{fig:inelMFP} shows the compiled data from calculations for higher energies and experimental data for lower energies.

For water electronic excitations and ionizations outweigh vibrational interactions and become the main inelastic channel above $\sim 14$~eV, where dielectric formalism calculations are in reasonably good agreement with experimental data for ice \cite{Michaud2003} (see Appendix \ref{app:inel_cross_sections}). For lower energies, vibrational excitations dominate. To account for them we have taken recommended CSs for a water molecule \cite{Itikawa2005} and added them to the calculated electronic CSs in order to obtain the total inelastic MFP. The result is the long-dashed line shown in Fig.~\ref{fig:inelMFP}, which agrees well with the experimental inelastic MFP for ice at low energies \cite{Michaud2003}.

For the coatings, since vibrational excitation data is not available, we have assumed that their CSs are the same as for water and we have followed the same procedure as for water.

A more in-depth analysis of the calculations, their benchmark, and the use of experimental data for low energies is provided in
Appendix~\ref{app:inel_cross_sections}.

\subsubsection{Elastic scattering mean free paths \label{sec:elastic}}

The calculation of elastic scattering CSs (and the corresponding MFPs) for compounds at intermediate and high energies is a relatively simple problem since atomic CSs can be used to describe the target.
This additivity approximation for elastic CSs might not hold for LEEs, where the electron wavelength becomes comparable to the interatomic distances \cite{Liljequist2008}. However, we demonstrate below that this is a reasonable estimate to obtain $\lambda_{\rm el}$.

The CSs for elastic scattering of electrons with atoms can be calculated by solving the Dirac equation in a central field (a method known as partial wave analysis) \cite{Salvat2003,Salvat2005}. This problem has been widely studied and there exist sources for obtaining these CSs, both differential and integral, e.g. the NIST database \cite{Jablonski2004} or the ELSEPA code \cite{Salvat2005}. In this work, we have used the latter source since it allows calculation of the CSs down to 10~eV and permits one to include different ingredients in the interaction potential which increases reliability of the calculations at low energies.

\begin{figure}[t]
    \begin{centering}
    \includegraphics[width=0.45\textwidth]{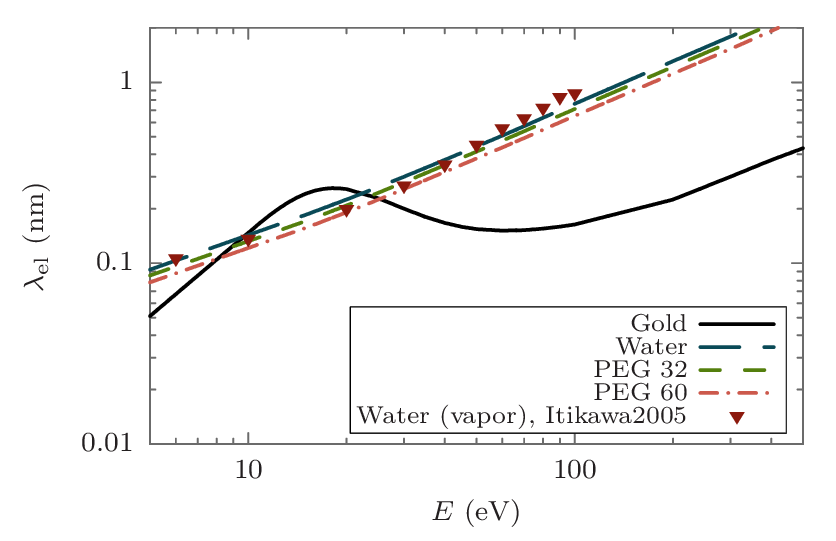}
    \par\end{centering}
    \caption{Elastic scattering mean free path $\lambda_\mathrm{el}$ of electrons in liquid water, gold and two PEG coatings. Lines represent calculations within relativistic partial wave analysis while symbols are recommended data for water (see text for details) \cite{Itikawa2005}.}
    \label{fig:elMFP}
\end{figure}

We have calculated the elastic scattering MFP by including the electrostatic, exchange, and polarization-corre\-lation potentials, as discussed in more detail in Appendix \ref{app:el_cross_sections}. The absorption potential was switched off since it accounts for inelastic processes which are already considered in the inelastic MFP. The results for gold, liquid water, and the PEG coatings are shown in Fig.~\ref{fig:elMFP} together with recommendations for water vapor (scaled to liquid water density) based on a compilation of extensive experimental and theoretical data \cite{Itikawa2005}. Data below 10~eV has been extrapolated from the calculated curves. The calculated values for water agree well with the recommended data even for energies below 10~eV. Therefore, the use of the atomic additivity rule for the elastic scattering CSs is justified. The MFPs $\lambda_{\rm el}$ for water and the PEG coatings are quite similar. Regarding gold, its elastic MFP presents a similar behavior at high energies, although being lower due to its higher density. At energies below about 50~eV $\lambda_{\rm el}$ for gold starts to grow, tending to converge with that for liquid water and the coating at the energies
of about $10 - 30$~eV.

The diffusion coefficients $D$ and average lifetimes $\tau$ based on the MFPs $\lambda_{\rm inel}$ and $\lambda_{\rm el}$ are presented in Table~\ref{tab:tauAndD}. Further details on calculations and their benchmark are given in Appendices~\ref{app:inel_cross_sections} and \ref{app:el_cross_sections}.

\begin{table}[htb!]
	\centering
	\caption{The diffusion coefficient $D$ (in nm$^2$\,fs$^{-1}$) and average lifetime $\tau$ (in fs) for 5~eV and 25~eV electrons in gold, the coating medium formed with 32 and 60 PEG ligands, and in pure water.}
	\label{tab:tauAndD}
	\begin{tabular}{llllll}
\hline
            \multicolumn{1}{c}{} &
            \multicolumn{1}{c}{} &
			\multicolumn{4}{c}	{Material}	\\	\cmidrule(l){3-6}

        Energy & Quantity & Gold & PEG 32	& PEG 60 & Water	 \\
\hline

	$\multirow{2}{*}{5~eV}$ 	& $D$	& 0.011	& 0.020	& 0.018 & 0.020	\\	
        	 						& $\tau$& 304	& 3.84 	& 3.84 & 3.84	\vspace{0.15cm} \\	

 	$\multirow{2}{*}{25~eV}$	& $D$	& 0.11	& 0.12  & 0.11	& 0.13 \\	 		  	
        	 						& $\tau$& 0.39	& 0.80	& 0.73	& 1.01 \\	
\hline
	\end{tabular}
\end{table}

\subsection{Evaluation of the number of emitted electrons \label{sec:pra}}

In this work it is assumed that LEEs are produced as a result of collective electron excitations arising in the metallic core of the NP \cite{Verkhovtsev2014a}. Two main contributions are taken into account, namely plasmon-type excitations of delocalized valence electrons and excitations of $5d$-electrons in individual atoms of gold.

The production of electrons via the plasmon excitation mechanism is quantified by relating the singly differential ionization cross section $\d\sigma_{\rm pl} / \d E$ (where $E$ stands for the electron kinetic energy)
to the probability of producing $N_\mathrm{e}^{\rm pl}$ electrons with energy within the
interval $\d E$, emitted from a segment $\d x$:
\begin{equation}
\frac{{\rm d}^2 N_\mathrm{e}^{\rm pl}}{{\rm d}x \,{\rm d}E} =
\frac{1}{V} \frac{ {\rm d}\sigma_{\rm pl} }{{\rm d}E} \ ,
\label{d2N_pl}
\end{equation}
where $V$ is the volume occupied by the metal core.
To calculate this quantity the cross section $\d\sigma_{\rm pl} / \d E$ is redefined as a function of the energy loss by the projectile, $\Delta\E = E + I_p$, with $I_p$ being the ionization threshold of the target.

The contribution of plasmon excitations is described by means of the plasmon resonance approximation (PRA)
\cite{Kreibig_Vollmer, Connerade_AS_PhysRevA.66.013207, Solovyov_review_2005_IntJModPhys.19.4143, Verkhovtsev_2012_EPJD.66.253}.
This methodology is briefly outlined below while a more detailed explanation can be found, e.g., in~\cite{Connerade_AS_PhysRevA.66.013207, Solovyov_review_2005_IntJModPhys.19.4143, Verkhovtsev_2012_EPJD.66.253, Verkhovtsev2014a}.

Within the PRA, the differential cross section $\d\sigma_{\rm pl} / \d\Delta\E$ for a spherical NP is defined as a sum of the surface (s) and the volume (v) plasmon terms,
\begin{equation}
\frac{\d\sigma_{\rm pl}}{\d\Delta\E} =
\frac{2\pi}{p_1 p_2} \int_{q_{\rm min}}^{q_{\rm max}} q \, \d q \,
 \left( \frac{\d^2\sigma_{\rm s}}{\d\Delta \E \, \d\Omega_{{\bf p}_2}}  +
 \frac{\d^2\sigma_{\rm v} }{\d\Delta \E \, \d\Omega_{{\bf p}_2}} \right)
\label{eq_01}
\end{equation}
which are constructed as a sum over different multipole contributions corresponding to different values of the angular momentum $l$.
Here,
$\Delta\E = \E_1 - \E_2$ is the energy loss of the incident projectile
of energy $\E_1$ while
${\bfp}_1$ and ${\bfp}_2$ are the initial and the final momenta of the projectile respectively,
${\bfq} = {\bfp}_1 - {\bfp}_2$ is the transferred momentum, and $\Om_{{\bfp}_2}$ is its solid angle.
Explicit expressions for the CSs entering Eq.~(\ref{eq_01}) are presented in Ref.~\cite{Verkhovtsev_2012_EPJD.66.253}.
They are obtained within the first Born approximation, which is applicable since the considered collision velocity ($v_{\rm ion} \approx 3.5$~a.u for a 0.3~MeV/u ion) is significantly larger than the characteristic velocity of delocalized electrons in gold NPs ($v_{\rm el} \approx 0.5$~a.u).

\begin{figure}[t]
    \begin{centering}
    \includegraphics[width=0.45\textwidth]{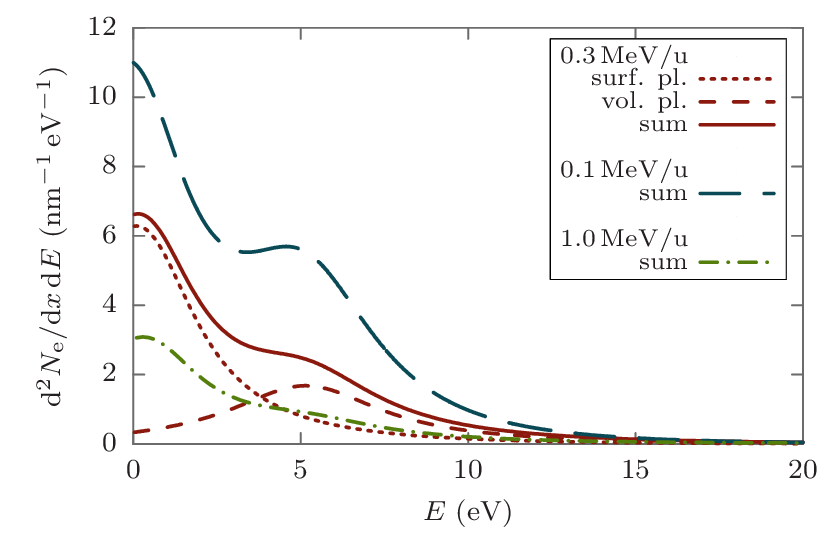}
    \par\end{centering}
    \caption{Number of electrons per unit length per unit energy emitted via the plasmon excitation mechanism from a AuNP of diameter 1.6~nm irradiated by carbon ions of different energy as indicated. For the 0.3~MeV/u case, dotted and dashed curves show the contributions of the surface and the volume plasmons respectively.}
    \label{fig:d2N_pl_energies}
\end{figure}

Figure~\ref{fig:d2N_pl_energies} shows the number of electrons per unit length per unit energy emitted via the plasmon excitation mechanism from a ``naked'' AuNP of diameter 1.6~nm irradiated with a 0.3~MeV/u carbon ion. The number of electrons produced under 0.1~MeV/u and 1.0~MeV/u ion irradiation is also shown for comparison.
The figure shows that the main contribution to the production of very LEEs comes from the surface plasmon (dotted red curve), while the volume plasmon (short-dashed red curve) contributes to the production of electrons of about 5~eV.
Our earlier analysis of LEE production by ``naked'' metallic NPs irradiated with $0.1 - 10$~MeV protons showed that the contribution of the surface plasmon mechanism becomes more prominent at higher impact energies and smaller NP size. For instance, in the case of NPs of 1~nm diameter the contribution of the surface plasmon to the ionization CS, and thus to the LEE yield, exceeds that of the volume plasmon by an order of magnitude \cite{Verkhovtsev2014a,Verkhovtsev2015a}.

The main contribution to the ionization CS due to plasmons comes from the dipole excitation mode when the characteristic collision distance between a projectile and a metallic core largely exceeds the radius of the core, $v_{\rm ion}/\Delta\varepsilon \gg R$ \cite{Solovyov_review_2005_IntJModPhys.19.4143, Gerchikov_1997_JPB.30.4133}. Here $v_{\rm ion}$ is the projectile velocity and $\Delta\varepsilon$ is a characteristic energy transferred during the collision.
At large impact parameters the dipole contribution dominates over those of the higher multipoles, since the dipole potential decreases slower at large distances than the higher multipole potentials.
For a 0.3~MeV/u ion (with velocity of $\sim 3.5$~a.u) and a characteristic energy transfer of about 0.2~a.u.~$ = 5$~eV, the characteristic impact parameter is approximately equal to 18~a.u.~$ = 0.9$~nm and is slightly larger than the core radius $R = 0.8$~nm.
The corresponding geometry of the collision is schematically depicted in Figure \ref{fig:outline}. In this case the dipole ($l = 1$), quadrupole ($l=2$) and octupole ($l=3$) plasmon excitation modes make comparable contributions to the ionization CS. The curves shown in Figure~\ref{fig:d2N_pl_energies} account for the contribution of the plasmon modes with $l = 1, 2$ and $3$.

Integration of (\ref{d2N_pl}) over the kinetic energy of emitted electrons gives the characteristic distance over which the projectile ion should traverse to ionize the NP via the plasmon excitation mechanism. For a 0.3~MeV/u ion this distance is about 0.03~nm. This means that when the ion has passed the 1.6~nm AuNP about 50 electrons are emitted due to the plasmon excitations formed in the metallic core. Decay of each plasmon excitation leads to emission of an electron; therefore, multiple plasmons are considered as independent processes taking place during the ion passage. For a 0.3~MeV/u ion ($v_{\rm ion} \approx 3.5$~a.u.) passing along the NP with the radius $R = 0.8$~nm~$ = 15.2$~a.u. the traversal will take the time $2R/v_{\rm ion} \approx 8.7$~a.u.~$ \approx 0.21$~fs, which is comparable to the lifetime of the surface plasmon resonance.
The lifetime of the plasmon resonance is defined as its inverse width, $\tau_{\rm pl} = 1/\Gamma = R/(3 l v_{\rm F})$, with $l$ being the angular momentum of the plasmon excitation and $v_{\rm F}$ the velocity of the NP valence electrons on the Fermi surface \cite{Gerchikov_2000_PRA.62.043201,Yannouleas_1992_AnnPhys.217.105}. For the dipole mode ($l = 1$) and for $v_{\rm F} = \sqrt{2 I_{p}} \approx 0.65$~a.u., an estimate for the lifetime $\tau_{\rm pl}$ gives 7.9~a.u.~$ = 0.19$~fs.
Therefore, even though a large number of electrons are emitted via the plasmon excitation mechanism, most of them will not escape the NP during the characteristic duration of the ion's passage. Valence electrons participating in the surface plasmon excitations are localized on the surface and emitted from there but the characteristic velocity of these electrons, $v \approx 0.4$~a.u., is an order of magnitude smaller than the velocity of the ion. This allows us to neglect the variation of the target charge state during ion passage and assume that the projectile ion interacts with a neutral target.

We have also accounted for electron emission via the collective electron excitations in individual atoms of a NP. The number of electrons $N_\mathrm{e}^{5d}$ with energy within the
interval $\d E$, emitted from a segment $\d x$ produced via the excitation of $5d$ electrons in individual gold atoms is defined as
\begin{equation}
\frac{{\rm d}^2 N_\mathrm{e}^{5d}}{{\rm d}x \,{\rm d}E} =
 \, n_t \frac{ {\rm d}\sigma_{5d} }{{\rm d}E} \ ,
\label{d2N_5d}
\end{equation}
where $n_t$ is the atomic number density of the target.

\begin{figure}[t]
    \begin{centering}
    \includegraphics[width=0.45\textwidth]{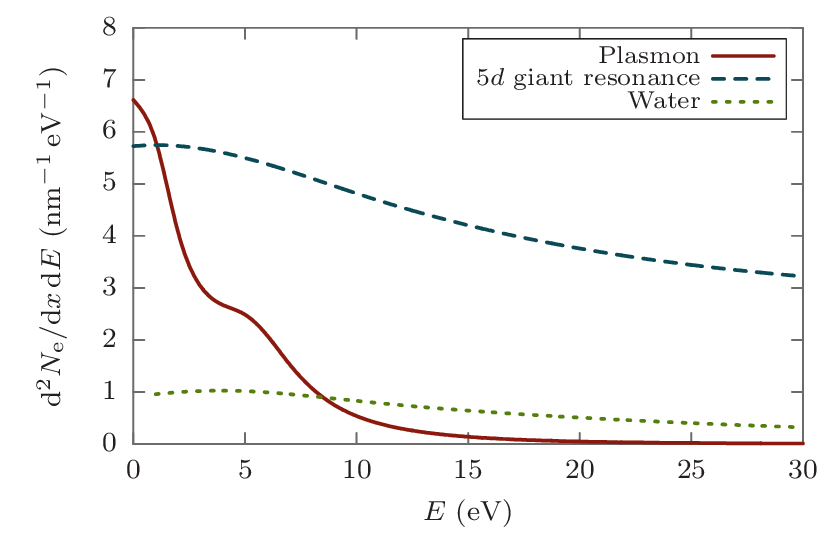}
    \par\end{centering}
    \caption{
    Number of electrons per unit length per unit energy produced via the plasmon (solid curve) and the $5d$ excitation mechanisms (dashed curve) in the 1.6~nm AuNP irradiated by a 0.3~MeV/u C$^{6+}$ ion. The dotted curve represents the number of electrons generated along the equivalent segment of ion track in water, calculated with the same methodology as outlined in Ref.~\cite{deVera2013}.
}
    \label{fig:electron-emission}
\end{figure}

The corresponding cross section $\d\sigma_{\rm 5d}/ \d\Delta\E$ (also redefined as a function of energy transfer $\Delta\E$) has been evaluated by means of an analytical expression which relates the cross section of photoionization with that of inelastic scattering in the dipole approximation~\cite{Verkhovtsev2014a,Verkhovtsev2015a}:
\begin{equation}
\frac{\d\sigma_{5\d}}{\d\Delta\E} =
\frac{2 Z^2 c}{\pi \Delta\E \, v_{\rm ion}^2} \,
\sigma_{\gamma}(\Delta\E)
\ln{ \left( \frac{v_{\rm ion}}{\Delta\E R_{5\d}} \right)} \ .
\label{ElScatter_to_PI}
\end{equation}
Here, $\sigma_{\gamma}(\Delta\E)$ is the photoionization cross section of the $5d$ electron shell in the atom of gold, $R_{5\d} \approx 2$~a.u is its characteristic radius and $v_{\rm ion}$ is the projectile velocity.
Equation~\eqref{ElScatter_to_PI}, obtained within the
``logarithmic approximation'', assumes that the main contribution to the cross section ${\rm d}\sigma_{\rm 5d} / {\rm d}\Delta\E$ comes from the impact parameter values in the range $R_{5\d} \lesssim r \lesssim v_{\rm ion}/\Delta\E$.
Considering a typical energy transfer associated with the $5d$ giant resonance, $\Delta\E \approx 25$~eV~$ \approx 1$~a.u \cite{Verkhovtsev2015a}, this estimate for a 0.3~MeV/u-ion results in $r \le 0.25$~nm. It suggests that the $5d$ collective electron excitations are formed in the atoms confined within a 0.25~nm-radius cylinder from the ion's path. This distance is smaller than the nearest-neighbor distance in bulk gold ($d = 0.288$~nm) and is comparable with that in small Au$_N$ ($N \le 20$) clusters \cite{Assadollahzadeh_2009_JCP.131.064306}. Therefore, only the atoms located close to the ion's path are excited via this mechanism for a 0.3~MeV/u-ion.
The number of electrons per unit distance per unit energy, ${\rm d}^2 N_\mathrm{e}^{5d}/{\rm d}x \,{\rm d}E$, was evaluated by averaging over different positions of the ion track with respect to the metal core, ranging from a central collision up to a glancing collision. This is done because the effect manifests itself stronger at small impact parameters $r \lesssim v_{\rm ion}/\Delta\E \approx 3.5$~a.u.
Our estimate shows that for the NP of diameter 1.6~nm which we consider as a case study approximately 15 gold atoms will be excited.

Figure~\ref{fig:electron-emission} shows the yield of electrons produced by a ``naked'' 1.6~nm AuNP irradiated by a 0.3~MeV/u C$^{6+}$ ion. The solid line shows the contribution of the plasmon excitations while the dash\-ed line presents the contribution from the atomic $5d$ giant resonance, evaluated using Eqs.~\eqref{d2N_pl} and~\eqref{d2N_5d}, respectively.
These results are compared with the number of electrons produced along the equivalent segment of ion track in water, calculated using the methodology of Ref.~\cite{deVera2013}.

\section{Results and discussion \label{sec:results}}

For further analysis we quantified the number of electrons emitted in the range $0 - 10$~eV which are produced by both collective electron excitation mechanisms as well as the number of electrons emitted in the range $10 - 30$~eV which are mainly due to the excitation of $5d$ electrons in individual atoms.
We approximate the diffusion of these electrons by considering them as two populations of electrons with characteristic energies 5~eV and 25~eV respectively. The corresponding diffusion coefficients $D$ and average lifetimes, $\tau$, obtained from the compiled curves shown in Figs.~\ref{fig:inelMFP} and \ref{fig:elMFP}, are summarized in Table~\ref{tab:tauAndD}.
The calculated number densities of first-generation electrons for both of these energies are presented in the following section.
In Section~\ref{sec:secondGen-results}, we present the results related to the second-generation of electrons resulting from the inelastic collisions of first-generation electrons with an energy of 25~eV. First-generation electrons of energy 5~eV are below the ionization potential of water and PEG, and thus are not capable of producing the second generation in our model.
In Section~\ref{sec:radical-production}, we analyze the radical production resulting from the decay of first-generation electrons taking into account the water content of the coating.
Finally, Section~\ref{NP_different_param} describes how variation of the projectile ion energy and the coating composition affects the production of electrons and OH radicals in the vicinity of the coated gold NPs.

\subsection{First-generation electrons}

\begin{figure}[t]
	\begin{centering}	
    \includegraphics[width=0.45\textwidth]{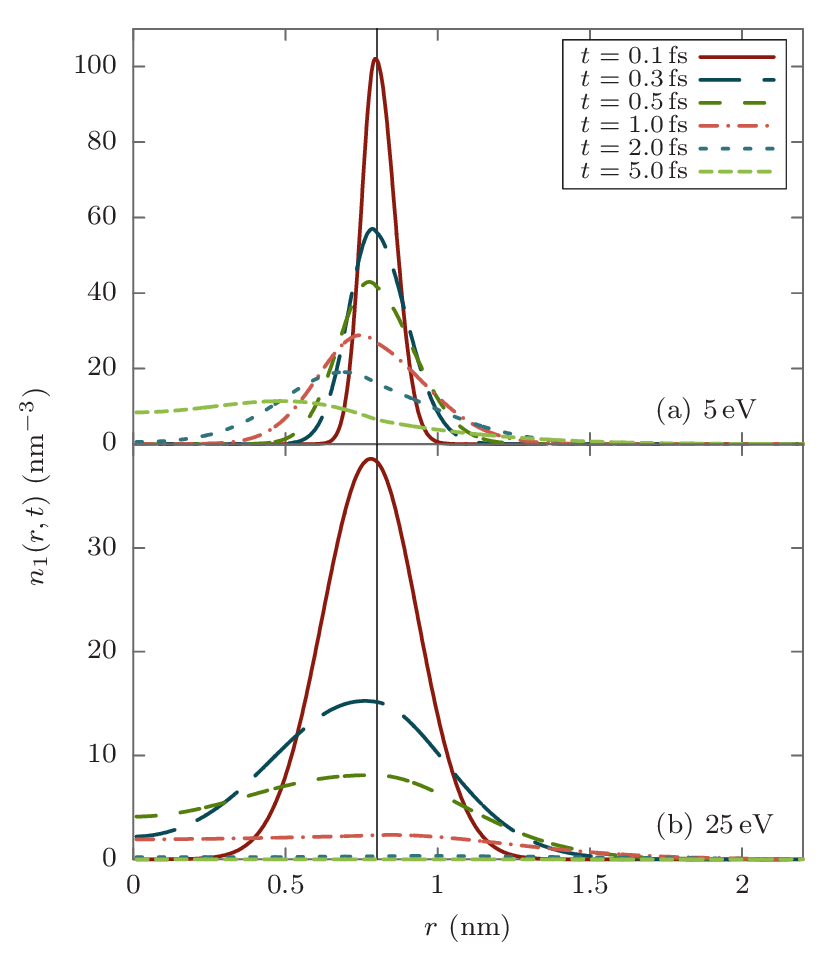}
	\par\end{centering}
    \caption{Number density of first-generation electrons $n_1(r,t)$ of energy (a) 5~eV and (b) 25~eV at various time instances $t$ versus radial distance from the NP center for the PEG~32 coating.
    The time instances indicated correspond to both panels. The vertical line at $r = 0.8$~nm illustrates the radius of the metal core.}
    \label{fig:n1o-peg32}
\end{figure}

The number density $n_1(r,t)$ of 5~eV and 25~eV first-genera\-tion electrons emitted from the AuNP surface for the case of $N_\mathrm{PEG} =32$ is shown in Fig.~\ref{fig:n1o-peg32} at various time instances.
As follows from the solution of Eq.~(\ref{Diff_eq_02}), the electrons diffuse away from the surface in both directions at a rate given by the diffusion coefficients $D_{\rm i}$ and $D_{\rm o}$.
The 25~eV-electrons have diffusion coefficients $D_{\rm i}$ and $D_{\rm o}$ about six and ten times larger respectively than the 5~eV-electrons which is evident by the faster broadening of the density.
At the same time, the number density of electrons diffusing away from the surface decreases exponentially due to inelastic collisions. A relative difference in $\tau$ for PEG~32 and PEG~60 of about a factor five leads to a significantly slower attenuation of the 5~eV-electrons compared to the 25~eV-electrons which have almost disappeared after 2~fs. This attenuation is responsible for the formation of second-generation electrons as discussed below.
The slope of number density $n_1(r,t)$ changes when crossing the surface of the NP core, which is clearly seen for 5~eV in the Fig.~\ref{fig:n1o-peg32}(a). This can be explained by a very large difference in the value of $\tau$ for gold and for PEG, see Table~\ref{tab:tauAndD}.

To quantify the number of electrons escaping the coating we calculated the fluence of electrons integrated over the area of a sphere with radius $r$ as a function of distance from the NP surface, as given by Eq.~\eqref{eq:1stgen-fluence}. The calculations were performed for the PEG~32 and PEG~60 coatings as well as for the ``naked'' AuNP (i.e., with no coating).
The integral fluence evaluated at the end of the coating for the two electron energies, 5~eV and 25~eV, is presented in Table~\ref{tab:1stGen-fluences}, where these numbers are normalized to the case of no coating.
In the case of the two coatings, electrons in the denser PEG 60 coating have a  lower diffusion coefficient which ultimately leads to about 35\% fewer electrons reaching the end of the coating compared to the PEG 32 coating.
For 5~eV-electrons, there is no difference between PEG 32 coating and no coating because the diffusion coefficient and average lifetime for the coating was approximated as identical to those of water.
For the 25~eV-electrons, however, there is a significant difference between the diffusion coefficient and average lifetime of electrons propagating through the coatings and in water. This leads to a reduction in the number of electrons escaping the coating of about 47\% and 65\% for the PEG~32 and PEG~60 coatings, respectively.
It should be stressed, however, that in all cases the vast majority of the emitted electrons experience an inelastic collision before reaching the coating end with just about 1\% of the electrons escaping the coating through purely elastic diffusion.

\begin{table}[htb!]
	\centering
	\caption{Integral fluence $F(r)$ of first-generation electrons evaluated at the end of the coating ($r - R = 1.4$~nm) for the two coatings PEG 32 and PEG 60 as well as with no coating. The fluence is normalized to the case of no coating.}
	\label{tab:1stGen-fluences}
	\begin{tabular}{llll}
\hline
            \multicolumn{1}{c}{} &
			\multicolumn{3}{c}	{Coating medium}	\\	\cmidrule(l){2-4}

    Energy  		& PEG 32& PEG 60& No coating\\
\hline
	5~eV 	& 1.0	& 0.65	& 1.0	\\
 	25~eV	& 0.53 & 0.35	& 1.0 \\
\hline
	\end{tabular}
\end{table}

\subsection{Second-generation electrons}
\label{sec:secondGen-results}

For electrons with an initial energy of 25~eV each inelastic collision leads to the production of two second-generation electrons, as discussed above in Section~\ref{sec:Methodology}. If we assume the mean ionization energy for the valence bands of the coating molecules of 15~eV then the remaining 10~eV after the ionizing collision are split evenly, resulting in the formation of two 5~eV second-generation electrons for each inelastic collision in the coating layer --- see Appendix~\ref{app:inel_cross_sections}.

\begin{figure}[t]
	\begin{centering}	
    \includegraphics[width=0.45\textwidth]{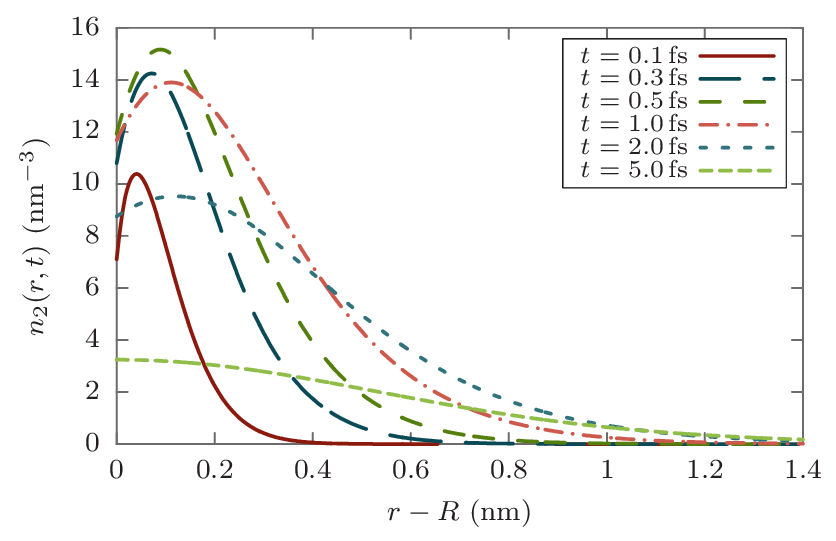}
	\par\end{centering}
	\caption{Number density of 5~eV second-generation electrons $n_2(r,t)$ in PEG~32 versus distance from the NP surface for various time instances.}
    \label{fig:n2o-peg32}
\end{figure}

The number density of second-generation electrons \linebreak $n_2(r,t)$ in the coating layer is described by Eq.~\eqref{eq:n2full} with $D$ and $\tau$ taken from Table~\ref{tab:tauAndD}. The corresponding results are plotted in Fig.~\ref{fig:n2o-peg32} at various time instances for the case of PEG~32.
Here we consider the contribution of the second-generation electrons that have been formed in the coating region and propagated outwards. Therefore Fig.~\ref{fig:n2o-peg32} shows the number density $n_2(r,t)$ as a function of radial distance from the NP surface.
As described in Section \ref{sec:diffusion} to accurately account for the formation of second-generation electrons within the metal core and their propagation to the outside region, one must derive a Green's function which takes into account the propagation of electrons in both the inside and the outside regions simultaneously.

\begin{figure}[t]
	\begin{centering}	
    \includegraphics[width=0.45\textwidth]{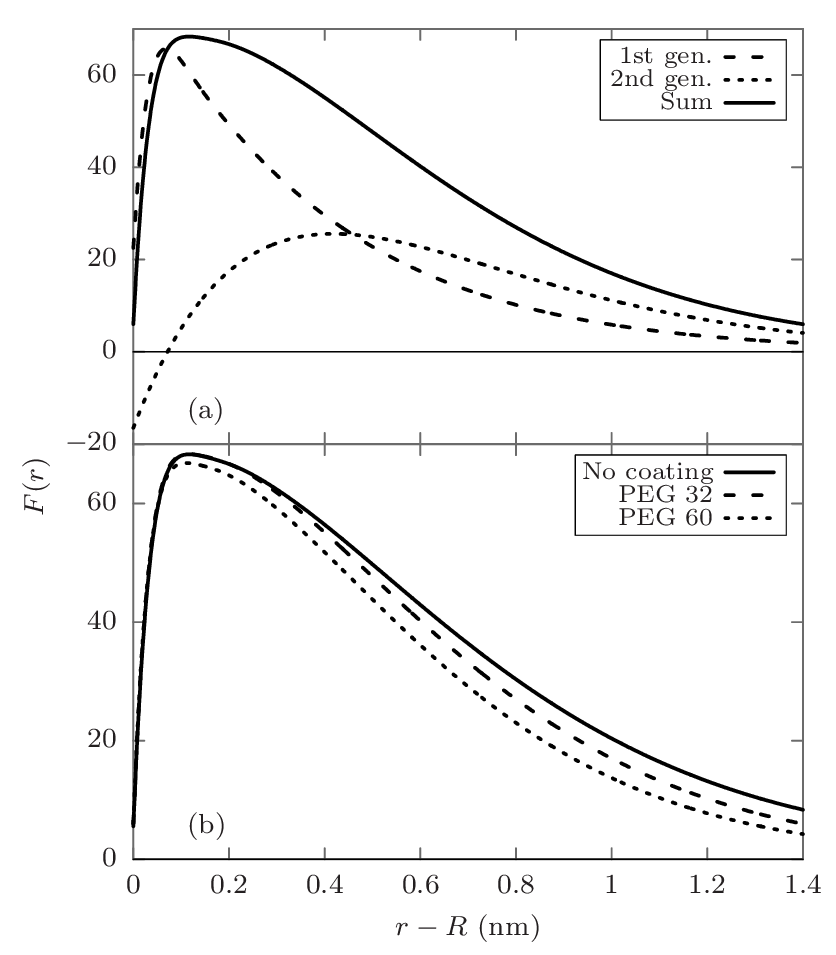}
	\par\end{centering}
	\caption{(a) Integral fluence $F(r)$ of the first and second-generation electrons and their sum in PEG~32. (b) Sums of integral fluence $F(r)$ due to first and second-generation electrons in PEG~32, PEG~60, and for no coating. }
    \label{fig:fluence2o-peg32}
\end{figure}

Figure~\ref{fig:n2o-peg32} illustrates that the density initially increases with time as first-generation electrons undergo inelastic collisions and decay into second-generation electrons. The maximum is reached after about 0.5~fs, after which the decay of second-generation electrons outweighs their generation. Unlike the first-generation electrons, those of the second generation can have their origin anywhere in the coating layer provided a first-generation electron attenuated at that point. This leads to a substantially more spread out number density profile for the second-generation electrons than that of the electrons emitted directly from the metal core.
The end result is a significant contribution to the integral fluence in the coating as can be seen in Fig.~\ref{fig:fluence2o-peg32}. It shows the fluence for the two generations as well as their sum for PEG~32 in the upper panel and the sums for PEG~32, PEG~60 and for no coating in the lower panel. The integral fluence at the coating edge, $r - R = 1.4$~nm, experiences around a three-fold increase when including the second-generation of electrons in the case of the two coatings.
The number density of second-generation electrons is maximum about 0.1~nm outside the NP core, as seen is Fig~\ref{fig:n2o-peg32}. This means that there is a flux of electrons towards the surface of the metal core which leads to a negative fluence for the second generation close to the core.
This is the reason why the sum of fluences of the first and second generation is slightly lower than that of the first generation alone.

\subsection{Production of hydroxyl radicals}
\label{sec:radical-production}

As mentioned previously, free radicals are responsible for the vast portion of biodamage produced by ionizing radiation. It is therefore necessary to accurately predict the number of radicals produced by secondary electrons emitted from a coated NP.

When electrons of sufficient energy emitted from the NP inelastically collide with the coating medium they will decay and produce two second-generation electrons, as discussed in Section~\ref{sec:Methodology}. If the colliding electrons have sufficient energy to electronically excite/ionize water molecules (i.e., when their energy is greater than the energy gap in water, $E_{\rm g} = 7$~eV), the excited/ionized water molecule may dissociate through different channels to produce an OH radical \cite{Nikjoo2006,Kreipl2009}.
In the present analysis this means that only first-generation electrons of 25~eV energy will result in radical formation.
LEEs with lower energies can also contribute to the OH radical production, e.g., through the process of dissociative electron attachment (DEA) \cite{Sanche_2005_EPJD.35.367}. This channel can, in principle, be included into the developed framework since the DEA cross sections for water can be taken from experiments or \textit{ab initio} calculations \cite{Itikawa2005,Fedor_2006_JPB.39.3935, Lacombe_2015_EPJD.69.195}. However, as expected from measurements of the cross section of DEA on water \cite{Itikawa2005}, the probability of attachment is very small ($< 10^{-4}$) \cite{Nano-IBCT2017}. Therefore, in the present analysis we neglect this contribution as a first approximation. The products, if any, of inelastic collisions of electrons with PEG molecules have not been studied and it is not part of the present analysis.

\begin{figure}[t]
	\begin{centering}	
    \includegraphics[width=0.45\textwidth]{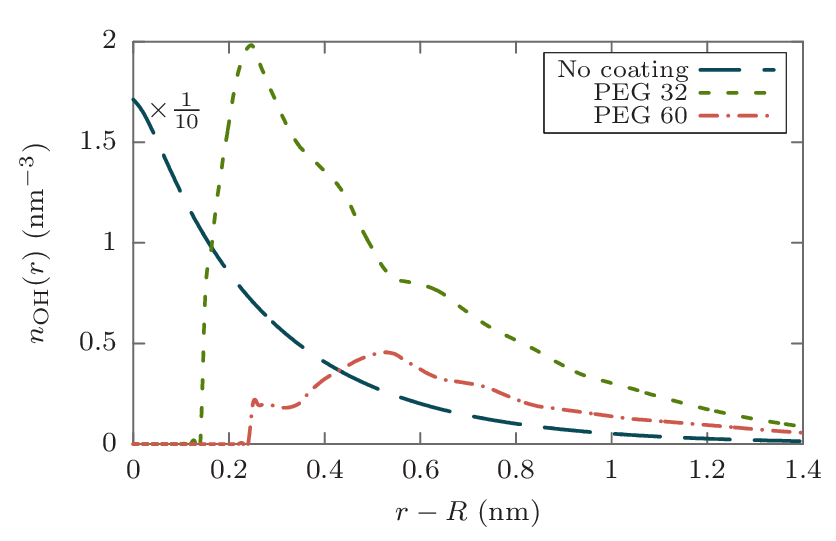}
	\par\end{centering}
	\caption{Number density of radicals $n_\mathrm{OH}(r)$ produced by inelastic decay of first-generation electrons of energy 25~eV in the coating region for coatings PEG~32 and PEG~60 as well as for no coating (``no coating'' line has been scaled by 1/10).}
    \label{fig:radicalplot}
\end{figure}

As can be seen in Fig.~\ref{fig:densityDistr}, the amount of water in the coating varies with distance from the NP boundary. We include this effect by a probability factor $\alpha(r)$ of producing an OH radical in an inelastic collision given by the relative mass ratio of water at the distance $r$ from the NP boundary, $\alpha(r) = \rho_\mathrm{H_2O}(r)/(\rho_\mathrm{H_2O}(r)+\rho_\mathrm{PEG}(r))$. The production rate of OH radicals is then given by
\begin{equation}
\frac{\partial n_\mathrm{OH}(r,t)}{\partial t} = \frac{n_{1,E>I_p}(r,t)}{\tau_\mathrm{1,o}} \alpha(r)
\label{eq:dn_OH/dt}
\end{equation}
and the radical number density at a given time instance is found by integrating Eq.~\eqref{eq:dn_OH/dt} over time
\begin{equation}
n_\mathrm{OH}(r,t) = \int_0^t \d t' \, \frac{n_{1,E>I_p}(r,t')}{\tau_\mathrm{1,o}} \alpha(r).
\label{eq:n_OH}
\end{equation}
The integration is carried out from zero to about 5~fs, which is the time instance at which all first-generation electrons will have decayed (see Fig.~\ref{fig:n1o-peg32}).
The number density of radicals $n_\mathrm{OH}(r,t)$ is practically invariant with time for the timescales concerned in this analysis because of the much lower diffusion coefficient of OH radicals compared with electrons \cite{Surdutovich2015}. We will therefore treat $n_\mathrm{OH}$ as dependent on $r$ only, $n_\mathrm{OH}(r,t) \equiv n_\mathrm{OH}(r)$.
In Fig.~\ref{fig:radicalplot}, the number density of radicals $n_\mathrm{OH}(r)$ is plotted for the two coatings PEG~32 and PEG~60 as well as for no coating.
The coating has a dramatic effect on the production of radicals, especially for the PEG~60 coating.
In both coatings water is completely absent in the first 0.3~nm from the surface of the NP while this is where most LEEs of the first-generation decay (compare Figs.~\ref{fig:densityDistr} and \ref{fig:n1o-peg32}), leading to a significantly reduced radical number density compared to the case of a ``naked'' NP.

To evaluate the total number of OH radicals $N_\mathrm{OH}^\mathrm{NP}$ produced due to secondary electrons emitted from the NP, the integral of Eq.~\eqref{eq:n_OH} over the extent of the coating region (that is, from the NP boundary $r = R$ to $r = R + l_\mathrm{coat}$, corresponding to the thickness of the coating layer $l_\mathrm{coat}$) is added to the integral fluence of electrons evaluated at the coating boundary $F(R + l_\mathrm{coat})$:
\begin{equation}
N_\mathrm{OH}^\mathrm{NP} = F(R + l_\mathrm{coat}) + \int \d r \, 4\pi r^2  n_\mathrm{OH}(r).
\label{eq:N_oh-NP}
\end{equation}
This can be done because all first-generation electrons of energy 25~eV escaping the coating will eventually decay and produce
a radical (here we assume that the coating region is surrounded by pure water).
The second term on the r.h.s. of (\ref{eq:N_oh-NP}) simply integrates the number density of radicals already produced inside the coating.

In addition to the radicals produced due the secondary electrons emitted from the NP surface, the ion will itself ionize the medium along its track, as given by the ionization CS for the medium.
Here we calculate the number of OH radicals produced by the primary ion in the coating medium using the ionization probabilities for water molecules
(evaluated following the method outlined in \cite{deVera2013}), due to the similarity of ionization CSs for the coating and for water.
The ionization of the medium by an ion results in hydrolysis of water molecules and emission of electrons which may inelastically collide with other water molecules and produce more radicals, similar to the process for those electrons emitted from the NP surface.
The difference here is that ionization caused by the ion directly produces OH radicals along with electrons as a result of hydrolysis.
To take into account the fact that the water content of the coating varies with radial distance from the NP surface, we calculated the mean distance of the ion from the NP surface as it passes by the NP and evaluated the relative water content along the track using the parameter $\alpha(r)$ as defined above.
As an illustration we consider the case when the ion passes by the metal core at a distance of 0.1~nm from its surface (see Figure~\ref{fig:outline}).
Given a coating thickness of 1.4~nm, one finds that the length of the ion's track through the coating region is $l = 4.0$~nm and that the average radial distance to the NP surface is $\bar{r} = 0.6$~nm.
The average relative water content for an ion passing through the two PEG coatings can then be evaluated at that distance as $\bar{\alpha}_\mathrm{PEG32} = 0.43$ and $\bar{\alpha}_\mathrm{PEG60} = 0.22$.
In the case of no coating, $\bar{\alpha} = 1.0$ that corresponds to pure water medium.
To compare with the production due to electrons emitted from the NP we consider only those electrons produced with an energy up to 30~eV.

The number of radicals produced along the track in the two cases is calculated as the sum of (i) the number of ionization events produced by the primary ion and (ii) those produced by the electrons which have sufficient energy to cause ionization, multiplied by the relative water content:
\begin{equation}
N_\mathrm{OH}^{\mathrm{track},i}
= l
	\int\limits_{I_p}^{30~eV} \d E \, \frac{\d^2 N_\mathrm{e}^{\rm track}}{\d x \d E}
\, \bar{\alpha}_i \,
\label{eq:N_oh-track}
\end{equation}
where $i$ denotes one of the coating media (PEG~32, PEG~60) or no coating.
In this case we do not solve the diffusion equation for the electrons emitted along the ion track but assume that the electrons with energy higher than $I_p$ produce OH radicals when decaying on average as given by $\bar{\alpha}_i$.

\begin{table}
	\centering
    \caption{Ratio of the number of radicals produced in the vicinity of coated (with 32 and 60 PEG molecules) and ``naked'' gold NPs of diameter 1.6~nm to that produced by ions of different energies traversing a similar distance in pure water. In the former case the radicals are produced due to inelastic decay of first-generation electrons ($E < 30$~eV) emitted from the NP and due to hydrolysis around the ion track.}
\label{tab:radical-yield}
\begin{tabular}{l l l l l}
\hline
	Ion energy		& PEG 32& PEG 60& No coating& Water\\
    	(MeV/u)		& & & & \\
\hline
0.3	& 0.63	& 0.29	& 1.69	& 1	\\
1.0	& 1.39	& 0.66	& 3.0	& 1 \\
5.0	& 6.5	& 3.2	& 20.0	& 1 \\
10.0& 9.0	& 4.1 	& 28.5	& 1 \\
\hline
\end{tabular}
\end{table}

The radical yields for the different coating media normalized to that of pure water medium are summarized in Table~\ref{tab:radical-yield}.
For a 0.3~MeV/u C$^{6+}$ ion irradiation the radical production is substantially reduced in the presence of a coating. In this case the ratio of number of radicals produced near the NP to that produced in pure water is smaller than one for both coatings studied. The denser PEG~60 coating produces less than a third of the radicals produced in pure water.
The main difference between the two coatings is that the increased number of PEG ligands in PEG~60 leads to a larger volume devoid of water molecules close to the NP surface.
In the case of no coating there is a 70\% enhancement of the radical yield produced by LEEs which illustrates the importance of the coating being permeable to water.

As seen in Fig.~\ref{fig:n1o-peg32}, most first-generation electrons decay within the first 0.4~nm outside the NP boundary, so to maximize radical yield it is critical that water should be present in this region.
A similar conclusion was made by Gilles and coworkers~\cite{Gilles2014} who experimentally demonstrated a decrease of OH radical yield with increasing coating density. According to those results, the OH radical yield plateaued at a six-fold decrease for PEG-coated AuNPs compared to ``naked'' AuNPs. The authors proposed that the number of PEG molecules in the coating, not the coating thickness, is the main factor for the radical yield decrease, in part due to a reduced water content for increasingly dense coatings.
Gilles {\it et al.} considered two PEG coatings of similar thickness (2.3~nm vs. 1.9~nm) and coating surface density (2.7~nm$^{-2}$ vs. 1.9~nm$^{-2}$) but with different molecular weight of the PEG molecules that made up the coating (1000~Da vs. 4000~Da). It was concluded that the four times greater density of PEG in the latter case is the cause of the suppressed radical production.
It should be stressed, however, that the experimental conditions in Ref.~\cite{Gilles2014} were different from the simulated ones in the present study:
The size of experimentally studied AuNP core was about 32~nm in diameter and the PEG molecules were between 4 and 16 times longer than those considered in our study. Apart from that, the experiments were performed with 17.5~keV X-rays irradiation while the present analysis is performed for ion irradiation. The fact that similar conclusions were reached may indicate that the effect of reduced radical yield for PEG-coated AuNPs has a general nature and is not restricted to any particular system size or radiation modality but depends on the water content of the coating. Our results emphasize also the significance of LEEs in the formation of the yield of OH radicals.

Finally, let us mention that emission of more energetic $\delta$-electrons (e.g., with energies of about 100~eV and higher) from a metal NP should lead to an additional production of radicals in its proximity. In this case one may expect an increase in the radical yield compared to pure water medium even at the Bragg peak energies.
This can be explained by the large difference of ionization CSs for gold and for water and hence by the higher probabilities of emission of more energetic electrons.
This difference can be estimated as the ratio of the energy-loss functions for gold and for water which is about 10 for the transferred energy of 100~eV and even larger for larger energy transfers.

Energetic $\delta$-electrons are characterized by large ranges in matter. For example, electrons with energy of 100~eV propagating in gold and in liquid water have a range of about 3~nm and 5~nm respectively, and the ranges for more energetic electrons are even longer \cite{Meesungnoen2002,Pianetta2009}. This means that $\delta$-electrons emitted from a few nanometer-size NPs to a great extend will penetrate the coating and produce OH radicals in the surrounding water medium. The above mentioned ranges in water are significantly larger than the coating thickness considered in this work $(l_{\rm coat} = 1.4$~nm).

Let us consider, as an estimate, that a typical $\delta$-electron can be represented as a 100-eV electron. Such an electron emitted from the core will cross the coating and will produce about 7 OH radicals by ionization of the water medium surrounding the NP until it becomes a solvated electron.
Since the number of $\delta$-electrons emitted from the core into the surrounding water medium is an order of magnitude larger than that produced in pure water,  there will be, at least, an order of magnitude increase in the radical yield around the NP when such more energetic electrons are taken into account. Here we include this estimate for completeness while a more elaborated analysis of the contribution of $\delta$-electrons will be performed in a future work.

\subsection{Production of electrons and radicals at different projectile energies and coating compositions \label{NP_different_param}}

As demonstrated in the previous section for irradiation with a 0.3~MeV/u C$^{6+}$ ion, the OH radical yield produced after the interaction of LEEs ($E < 30$~eV) with water molecules for the two considered coatings, PEG~32 and PEG~60, is smaller than that in pure water. In this section we explore how the variation of coating composition and the kinematics of ion impact influence the production of LEEs and free radicals and thus the ability of the AuNP to act as a radiosensitizing agent.

\begin{figure}[t]
    \begin{centering}
    \includegraphics[width=0.45\textwidth]{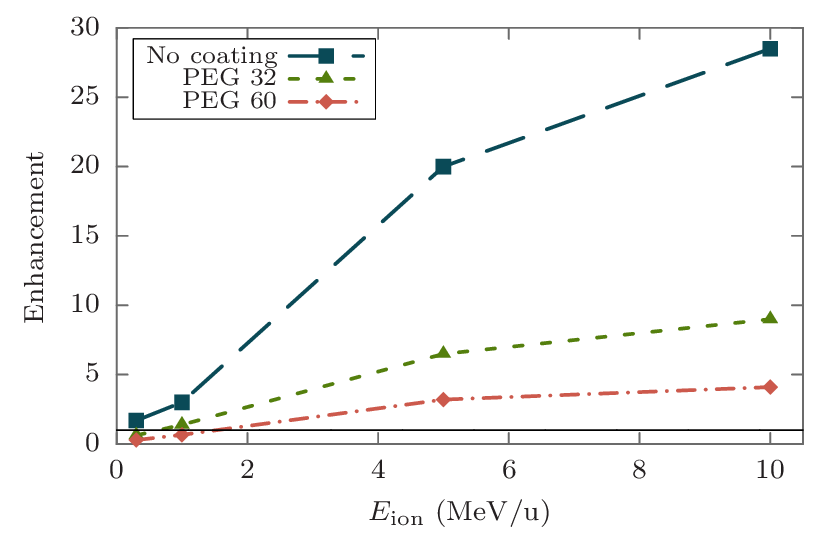}
    \par\end{centering}
    \caption{Relative increase of the radical yield (as compared to pure water) versus the projectile ion energy for PEG~32 and PEG~60 coating medium as well as for no coating. The horizontal line shows the case when the radical yield around the NP is equal to that in pure water. Lines connecting the points are meant only to guide the eye.}
    \label{fig:radical-yield-enhancement-energies}
\end{figure}

We start by analyzing the radical yield (relative to that in pure water) for different ion energies in the range 0.3 to 10~MeV/u. These results are presented in Table~\ref{tab:radical-yield} and shown in Fig.~\ref{fig:radical-yield-enhancement-energies}.
It was shown in Ref.~\cite{Verkhovtsev2014a} that at large collision velocities the ionization CS  of gold clusters due to plasmon excitations and hence their contribution to LEE production decrease.
The cross section of ionization due to $5d$ excitations, Eq.~(\ref{ElScatter_to_PI}), is maximum at the projectile energy or about 0.3~MeV/u and decreases at higher energies due to the $1/v_{\rm ion}^2$ term.
The ionization cross section for water behaves similarly. As a result, when a NP is irradiated by an ion with an energy exceeding that in the Bragg peak region in water, the number of electrons emitted from the NP will be larger than that produced by the same ion traversing water medium. This leads to a significant increase in the number of OH radicals generated around the NP as compared with pure water (see Fig.~\ref{fig:radical-yield-enhancement-energies}).

\begin{figure}[t]
    \begin{centering}
    \includegraphics[width=0.4545\textwidth]{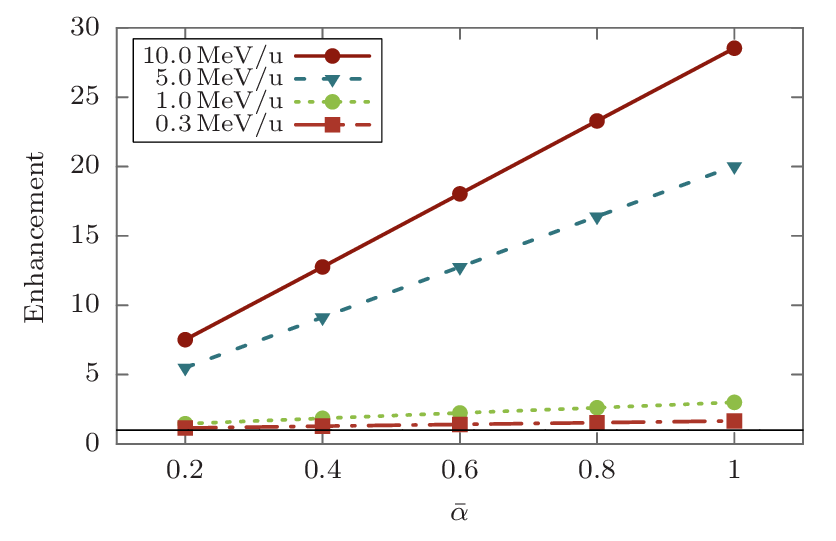}
    \par\end{centering}
    \caption{Relative increase of the radical yield versus average water content of the coating region assuming a homogeneous material calculated for different ion energies. The horizontal line shows the case when the radical yield around the NP is equal to that in pure water.}
    \label{fig:radical-yield-enhancement-water}
\end{figure}

Next, we illustrate the effect of the water content within the coating. The coating is approximated as a homogeneous material with an average relative water content $\bar{\alpha}$ defined in the same way as above.
We calculated the radical production in the coating region using Eqs. \eqref{eq:n_OH} and~\eqref{eq:N_oh-NP} but substituted $\alpha(r)$ with an average relative water content $\bar{\alpha}$. In this case we used the number density $n_1(r,t)$ calculated for the case of no coating. This assumption should not impact the results since the number density profiles are similar for all the studied coating media.
To this we added the contribution from radicals produced due to ionization along the ion track calculated using Eq.~\eqref{eq:N_oh-track} while $\alpha_i$ was substituted also with $\bar{\alpha}$.
The results are shown in Fig.~\ref{fig:radical-yield-enhancement-water}.
Approximating the coating region as a homogeneous material with an average water content leads to an enhanced production of radicals as compared to pure water, due to the presence of water molecules at the NP surface which are lacking when considering the realistic coating.

For the ion energy corresponding to the Bragg peak in water, the increase in the radical yield varies from 1.2 to 1.7 when the water content varies from 0.2 to 1.0. The moderate (20\% to 70\%) increase of the radical yield reflects that the fact that the production of radicals in pure water is high at such ion energies.
At higher energies the radical yield drastically increases as illustrated in Fig.~\ref{fig:radical-yield-enhancement-water}. For a 10~MeV/u ion the enhancement factor is 7.5 even for an average water content $\bar{\alpha} = 0.2$. This clearly demonstrates that for an efficient OH radical production due to LEEs emitted from the NP, having water present at the surface of the NP is more important than the total average water content. At higher ion energies there is a 4-fold increase in the radical yield due to an increase of the average water content. Thus the coating composition and especially its permeability to water molecules should be carefully considered when designing coated radiosensitizing NPs.

\section{Conclusion \label{sec:conclusion}}

In this work we have presented a coherent theoretical framework for studying the transport of secondary electrons emitted from coat\-ed metallic nanoparticles (NPs) embedded in water medium and irradiated by ions. Knowing the fluence of emitted electrons allows the yield of OH radicals created around the NP to be calculated.

Several theoretical and computational methods have been combined into a single framework: (i) the number of electrons emitted from the metallic core of a NP due to collective electron excitations as a result of ion irradiation was evaluated analytically by means of the plasmon resonance approximation;
(ii) information on the NP core and the coating structure was obtained from molecular dynamics simulations, (iii) the elastic and inelastic scattering cross sections between electrons and the coating material were evaluated by means of relativistic partial-wave analysis and the dielectric formalism, and finally
(iv) the transport of electrons emitted from the NP core through the coating was described by means of a diffusion equation-based approach.

The methodology presented in this work is general and can be applied for NPs of different size and coating thickness as well as for different chemical compositions of a core and a coating. It allows the yield of OH radicals around the NP to be evaluated and the radical yield to be analyzed as a function of projectile ion energy and at different parameters of the coating. Outcomes of such analysis can be compared with experimental data and results of Monte Carlo simulations. The present formalism complements the Monte Carlo approach by accounting for the important low-energy and many-body phenomena such as collective electron excitations in the metallic core.

As a case study we applied the developed formalism to study OH radical production around an Au@PEG NP with the core diameter of 1.6~nm and the coating thickness of 1.4~nm.
We found that nearly every low-energy electron emitted from the NP core due to collective electron excitations undergoes an inelastic collision in the coating region. Due to a low content of water molecules near the NP surface, the number of radicals produced in the coating due to these inelastic collisions is significantly smaller as compared to a ``naked'' AuNP of the same size.

We also analyzed the OH radical yield as a function of ion impact energy and as a function of the average water content in the coating region.
It was shown that the yields of low-energy electrons and free radicals at ion energies of 5 and 10~MeV/u are significantly larger than those in the Bragg peak region in water (0.3~MeV/u).

It was demonstrated that the presence of water at the NP surface is an important factor for efficient radical yield enhancement.
To maximize the radical yield due to LEE emission from coated metal NPs, it is therefore recommended to apply the least dense and/or most hydrophilic coating possible thereby allowing an increased water density at the NP surface.

A more systematic analysis exploring NPs and ions in a wider parameter space and including the production of more energetic electrons will be continued in our future work with the aim to explore further physical mechanisms underlying radiosensitization by NPs for therapeutic applications with ion beams. We expect the present model to be of great importance for this purpose, owing to the generality of the developed methodology.

\section*{Author contribution statement}
All authors contributed to the development of the methodology as well as to analysis and interpretation of the results.
KH performed MD simulations of nanoparticle structure, PdV calculated electron interaction cross sections,
and AV calculated electron emission cross sections from the metallic core.
KH, PdV and AV drafted the manuscript with input from ES, NJM and AVS.
All authors discussed the results and contributed to the final manuscript.

\bigskip{}

The research leading to these results has received funding from the European Union Seventh Framework Programme PEOPLE - 2013 - ITN - ARGENT project under grant agreement number 608163. PdV would like to thank Prof. Isabel Abril and Prof. Rafael Garcia-Molina for their help with the calculation of electron inelastic cross sections. AV is grateful to Dr. Andrei Korol for fruitful discussions.

\appendix
\section{Calculations of the electron number density \label{Appendix_Diff_eq}}
\subsection{Analytic solution of the diffusion equation \label{Appendix_Diff_eq_analytic}}

Let us consider the diffusion equation for a spherically symmetric source:
\begin{equation}
\frac{\partial n(r,t)}{\partial t} = D \, \frac{1}{r^2}
\frac{\partial}{\partial r} \left( r^2 \frac{\partial n(r,t)}{\partial r} \right)
- \frac{n(r,t)}{\tau}  \ ,
\label{AppA_01}
\end{equation}
and seek the solution of this equation in the form
\begin{equation}
n(r,t) = \tilde{n}(r,t) \, e^{-t/\tau} \ .
\end{equation}
Substituting this solution into Eq.~\eqref{AppA_01} one gets
\begin{equation}
\frac{\partial \tilde{n}(r,t)}{\partial t}
=
D \, \frac{1}{r^2}
\frac{\partial}{\partial r} \left( r^2 \frac{\partial \tilde{n}(r,t)}{\partial r} \right) \ .
\label{AppA_01_new}
\end{equation}
Let us now present the desired number density as $\tilde{n}(r,t) = \xi(r,t)/r$.
After this substitution Eq.~\eqref{AppA_01_new} transforms into
\begin{equation}
\frac{\partial \xi(r,t)}{\partial t} = D \, \frac{\partial^2 \xi(r,t)}{\partial r^2} \ ,
\label{AppA_02}
\end{equation}
which we can solve by representing $\xi(r,t)$ as an inverse Laplace transform defined as
\begin{equation}
\xi(r,t) = \frac{1}{2\pi i} \int_{\sigma - i\infty}^{\sigma + i\infty}
\d s \, e^{st} \, \tilde{\xi}(r,s) \ .
\label{AppA_03}
\end{equation}
The partial differential equation~\eqref{AppA_02} can then be transformed into an ordinary differential equation
\begin{equation}
s \, \tilde{\xi}(r,s) = D \, \frac{\partial^2 \tilde{\xi}(r,s)}{\partial r^2} \ ,
\end{equation}
whose solution reads as
\begin{equation}
\tilde{\xi}(r,s) = C_1 \, e^{- \sqrt{s/D} \, r} + C_2 \, e^{\sqrt{s/D} \, r} \ ,
\label{xi_r_s}
\end{equation}
where the constants $C_1$ and $C_2$ should be found using the initial and boundary conditions at the surface of the metallic sphere.
Substituting Eq.~\eqref{xi_r_s} into Eq.~\eqref{AppA_03} yields
\begin{multline}
\xi(r,t) =
\frac{C_1}{2\pi i} \int_{\sigma - i\infty}^{\sigma + i\infty}
\d s \, e^{st - \sqrt{s/D} \, r}  \\
+ \frac{C_2}{2\pi i} \int_{\sigma - i\infty}^{\sigma + i\infty}
\d s \, e^{st + \sqrt{s/D} \, r}  \ .
\end{multline}

Now let us take into account that the electrons emitted from the surface with the radius $R$ can travel in both the inner (``$<$'') and the outer (``$>$'') regions. Then,
\begin{align}
\xi(r,t) =
&\frac{1}{2\pi i} \int_{\sigma - i\infty}^{\sigma + i\infty}
\d s \, C_<^-(s) e^{st - \sqrt{s/D} \, r} \, H(R-r) \nonumber \\
&+
\frac{1}{2\pi i} \int_{\sigma - i\infty}^{\sigma + i\infty}
\d s \, C_<^+(s) e^{st + \sqrt{s/D} \, r} \, H(R-r) \nonumber \\
&+
\frac{1}{2\pi i} \int_{\sigma - i\infty}^{\sigma + i\infty}
\d s \, C_>^-(s) e^{st - \sqrt{s/D} \, r} \, H(r-R) \nonumber \\
&+
\frac{1}{2\pi i} \int_{\sigma - i\infty}^{\sigma + i\infty}
\d s \, C_>^+(s) e^{st + \sqrt{s/D} \, r} \, H(r-R) \ ,
\label{xi_general}
\end{align}
where $H(x)$ is the Heaviside step function.
The first two terms on the r.h.s. of Eq.~\eqref{xi_general} describe the part of the space occupied by the metallic core ($0<r<R$) while the latter two terms describe the outer space ($r>R$).
The constant $C_>^+(s)$ in the latter term on the r.h.s. can immediately be set to zero in order to avoid an exponential growth of $\xi(r,t)$ as $r \to \infty$.

The number density $\tilde{n}(r,t)$, as well as the function $\xi(r,t)$, should be continuous functions at the radius of the metallic core $r=R$:
\begin{align}
\tilde{n}(r,t)_{r \to R_-} &= \tilde{n}(r,t)_{r \to R_+} \ , \\
\xi(r,t)_{r \to R_-} &= \xi(r,t)_{r \to R_+} \nonumber \ ,
\end{align}
which implies the following relation
\begin{equation}
 C_<^-(s) e^{-\sqrt{s/D} \, R} + C_<^+(s) e^{\sqrt{s/D} \, R} =
 C_>^-(s) e^{-\sqrt{s/D} \, R} \nonumber \ .
\end{equation}
After performing the following substitutions
\begin{align}
C_<^-(s) &=  \tilde{C}_<^-(s) \, e^{-\sqrt{s/D} \, R} \ , \nonumber \\
C_<^+(s) &=  \tilde{C}_<^+(s) \, e^{-\sqrt{s/D} \, R} \ , \nonumber \\
C_>^-(s) &=  \tilde{C}_>^-(s) \, e^{\sqrt{s/D} \, R}
\end{align}
and carrying out some algebraic transformation the following relationship is derived
\begin{equation}
\tilde{C}_>^-(s) = \tilde{C}_<^-(s) \left[ e^{- 2 \sqrt{s/D} \, R} - 1 \right] \ .
\end{equation}

In what follows, we denote $\tilde{C}_<^-(s) \equiv C(s)$ for the sake of simplicity.
Then, Eq.~\eqref{xi_general} for $\xi(r,t)$ transforms into
\begin{align}
\xi(r,t) &=
\frac{1}{2\pi i} \int_{\sigma - i\infty}^{\sigma + i\infty}
\d s \, C(s) \, H(R-r)  \, e^{st} \nonumber \\
&\quad\times \left( e^{- \sqrt{s/D} \, (r+R)} - e^{\sqrt{s/D} \, (r-R)} \right) \nonumber \\
&\quad +\frac{1}{2\pi i} \int_{\sigma - i\infty}^{\sigma + i\infty}
\d s \, C(s) \, H(r-R)  \, e^{st} \nonumber \\
&\quad\times \left( e^{- \sqrt{s/D} \, (r+R)} - e^{- \sqrt{s/D} \, (r-R)} \right) \ ,
\end{align}
which can be further transformed to
\begin{align}
\xi(r,t) &= \frac{1}{2\pi i} \int_{\sigma - i\infty}^{\sigma + i\infty}
\d s \, C(s) \, e^{st} \nonumber \\
&\quad\times \left( e^{- \sqrt{s/D} \, (r+R)} - e^{- \sqrt{s/D} \, |r-R|} \right) \ .
\end{align}
The constant $C(s)$ should be negative to assure that the function $\xi(r,t) > 0$.
Let us then redefine this constant, $C(s) \equiv - C(s) > 0$, so that
\begin{multline}
\xi(r,t) =
\frac{1}{2\pi i} \int_{\sigma - i\infty}^{\sigma + i\infty}
\d s \, C(s) \, e^{st} \\
\times \left( e^{- \sqrt{s/D} \, |r-R|} - e^{- \sqrt{s/D} \, (r+R)} \right) \ ,
\end{multline}
and $C(s) > 0$ for any $r > 0$. This constant can be determined from the initial condition that all $N_\mathrm{e}$ electrons are emitted simultaneously at the time instance $t=0$:
\begin{align}
N_\mathrm{e} H(t) &=
\int_0^{\infty} \tilde{n}(r,t) \, 4 \pi r^2 \, \d r =
\int_0^{\infty} \xi(r,t) \, 4 \pi r \, \d r \nonumber \\
&=
\frac{1}{2\pi i} \int_{\sigma - i\infty}^{\sigma + i\infty}
\d s \, C(s) \, e^{st}  \nonumber \\
&\quad\times
4 \pi \int_0^{\infty} \d r \, r
\left( e^{- \sqrt{s/D} \, |r-R|} - e^{- \sqrt{s/D} \, (r+R)} \right) \ .
\end{align}
To calculate the latter integral, let us present it as a sum of two integrals, $\int_0^R$ and $\int_R^{\infty}$, because each of them can be solved analytically. After carrying out some algebraic transformations, one obtains
\begin{equation}
N_\mathrm{e} H(t) =
\frac{4 R \sqrt{D}}{i} \int_{\sigma - i\infty}^{\sigma + i\infty}
\d s \, e^{st} \, \frac{C_0}{s} \ ,
\end{equation}
where the substitution $C_0 = \sqrt{s} \, C(s)$ has been made. The constant $C_0$ can be calculated by carrying out a contour integration in the complex plane so that the point of singularity, $s = 0$, lies outside the contour. As a result, one gets
\begin{equation}
C_0 = \frac{N_\mathrm{e}}{8 \pi  R \sqrt{D}} \ ,
\end{equation}
and the expression for $\xi(r,t)$ transforms into
\begin{multline}
\xi(r,t) =
\frac{N_\mathrm{e}}{16 \pi^2 i R \sqrt{D}}
\int_{\sigma - i\infty}^{\sigma + i\infty}
\d s \, \frac{e^{st}}{\sqrt{s}} \\
\times \left( e^{- \sqrt{s/D} \, |r-R|} - e^{- \sqrt{s/D} \, (r+R)} \right) \ .
\label{eq_xi}
\end{multline}
The integration of the r.h.s. of (\ref{eq_xi}) gives $\xi(r,t)$ and, finally, $\tilde{n}(r,t) = \xi(r,t)/r$ is given by,
\begin{align}
\tilde{n}(r,t) = \frac{N_\mathrm{e}}{8 \pi^{3/2} R \, r \sqrt{Dt}}
\left[ e^{- (r-R)^2/4Dt} - e^{-(r+R)^2/4Dt} \right] .
\end{align}
This is the solution of the diffusion equation given by Eq.~\eqref{AppA_01_new}. Equation~\eqref{number_dens_1st_gen} is then obtained by multiplying this solution by an exponential factor $\exp({-t/\tau})$ due to attenuation of the emitted electrons.

\subsection{Accounting for the two different media \label{Appendix_Diff_eq_2media}}

Let us now assume that the electrons propagate inside the metallic core of the NP ($0 < r < R$) and in the outer region ($r \geq R$) with the diffusion coefficients $D_{\rm i}$ and $D_{\rm o}$ respectively and that the corresponding electron lifetimes are given by $\tau_{\rm i}$ and $\tau_{\rm o}$.
One can write then the inverse Laplace transform for the each region as
\begin{align}
\tilde{\xi}_\myin(r,s)
& = C_{1\myin} \, e^{ - \sqrt{(s + \gamma_\myin)/D_\myin} \, (r - R)}  \nonumber\\
& \quad + C_{2\myin} \, e^{\sqrt{(s + \gamma_\myin)/D_\myin} \, (r - R)} \ , & &0 < r < R   \nonumber \\
\tilde{\xi}_\myout(r,s)
& = C_{1\myout} \, e^{ - \sqrt{(s + \gamma_\myout)/D_\myout} \, (r - R)}  \nonumber \\
& \quad + C_{2\myout} \, e^{\sqrt{(s + \gamma_\myout)/D_\myout} \, (r - R)} \ , & &r \geq R
\end{align}
where $\gamma_{\myin,\myout} = 1/\tau_{\myin,\myout}$, respectively.

Accounting for the boundary conditions the following expressions are obtained:
\begin{align}
\tilde{\xi}_\myin(r,s) &= C_{2\myin} \, \left[
	e^{ \sqrt{(s + \gamma_\myin)/D_\myin} \, (r - R) } \right. \nonumber \\
	&\quad - \left. e^{ -\sqrt{(s + \gamma_\myin)/D_\myin} \, (r + R) }
	\right] & & 0 < r < R  \nonumber  \\
\tilde{\xi}_\myout(r,s) &= C_{2\myin} \, \left[
e^{ -\sqrt{(s + \gamma_\myout)/D_\myout} \, (r - R) } \right. \nonumber \\
&\quad - \left.
e^{ -2\sqrt{(s + \gamma_\myin)/D_\myin} \, R - \sqrt{(s + \gamma_\myout)/D_\myout}\, (r - R) }
\right] & & r \geq R \ .
\label{eq:xii-xio}
\end{align}
The coefficient C$_{2i}$ can be found from the following normalization condition which explicitly includes that some fraction of electrons has been attenuated:
\begin{multline}
\int_0^R 4 \pi r \,
\left[ 1 + \frac{\gamma_\myin}{s} \right] \tilde{\xi}_\myin(r,s) \, {\rm d}r \\
+ \int_R^{\infty} 4 \pi r \,
\left[ 1 + \frac{\gamma_\myout}{s} \right] \tilde{\xi}_\myout(r,s) \, {\rm d}r  = \frac{N_\mathrm{e}}{s} \ .
\end{multline}
If we denote the first integral as $I_\myin$ and the second integral as $I_\myout$, we then have the equation
\begin{equation}
C_{2\myin} \left( I_\myin + I_\myout \right) = \frac{N_\mathrm{e}}{s} \ .
\end{equation}
The solution of this equation gives us an expression for $C_{2\myin}$:
\begin{equation}
C_{2\myin} = \frac{N_\mathrm{e}}{s (A + B)} \ ,
\end{equation}
where
\begin{align}
A &= \frac{4\pi}{s} \,
\left[ 1 - e^{-2R \sqrt{(s + \gamma_\myin)/D_\myin} } \right]
\left( R \sqrt{D_\myout (s + \gamma_\myout)} + D_\myout \right)  \ ,
\nonumber \\
B &= \frac{4\pi D_\myin}{s} \,
\left[ R \sqrt{\frac{s + \gamma_\myin}{D_\myin}} - 1 \right. \nonumber \\
&\quad+
\left.
e^{- 2R \sqrt{(s + \gamma_\myin)/D_\myin} }
\left( R \sqrt{\frac{s + \gamma_\myin}{D_\myin}} + 1 \right)
\right] \ .
\end{align}

Setting $\gamma_\myin = \gamma_\myout$ and $D_\myin = D_\myout$, i.e. considering the diffusion in a homogeneous medium, allows for an analytical inverse Laplace transform of Eq.~\eqref{eq:xii-xio} and yields the same shape as the analytical result for the homogeneous case.

In the general, inhomogeneous case, one has to do the inverse Laplace transform numerically. This was done with \textsc{Mathematica} using the Fixed-Talbot algorithm\footnote{A \textsc{Mathematica} package is available at \url{http://library.wolfram.com/infocenter/MathSource/5026/}} as described in Ref.~\cite{InvLaplace}.

\section{Calculation of electron interaction mean free paths \label{app:cross_sections}}
\subsection{Dielectric formalism calculations for inelastic cross sections\label{app:inel_cross_sections}}

For electrons with sufficient energy the dielectric formalism can be used to obtain the relevant electronic interaction quantities \cite{Lindhard1954,Ritchie1977}. The inverse inelastic mean free path (IMFP) can be calculated as \cite{Abril2013}:
\begin{align}
\Lambda(E) &= \frac{1}{\lambda_{\rm inel}(E)} \label{eq:IIMFP} \\
&= \int_{\Delta \varepsilon_-}^{\Delta \varepsilon_+} \int_{q_-}^{q_+}
f_{\rm ex} \, \frac{e^{2}}{\pi \hbar} \frac{m}{E} \frac{1}{q}
  \, \mathrm{Im} \left[
\frac{-1}{\epsilon(\Delta \varepsilon, q)} \right] \d q \, \d \Delta \varepsilon {\rm ,} \nonumber
\end{align}
where the integration limits are set by energy and momentum conservation \cite{Garcia-Molina2016,deVeraThesis}. In the above equation $E$ is the electron kinetic energy, $\hbar$ is the reduced Planck's constant, and $e$ is the fundamental charge. The exchange factor $f_{\rm ex}$ accounts for the indistinguishability of the incident and the emitted electrons and usually has a noticeable influence for energies below about 300~eV. The target is characterized by its energy-loss function (ELF), ${\rm Im}[-1/\epsilon(\Delta \varepsilon, q)]$, related to the complex dielectric function $\epsilon(\Delta \varepsilon, q)$ which represents its electronic excitation spectrum.

The optical ELF ($q=0$) of liquid water and gold are taken from experimental data \cite{Hayashi2000,Palik1999} while the optical ELF of the PEG coating was calculated by using an empirical parameterization for organic materials \cite{Tan2004}. Based on the available experimental data for organic targets and realizing the similarities between them (that is, a main single excitation around 20~eV), Tan \textit{et al.} suggested the representation of an ELF by a single Drude-type function \cite{Tan2004}. Its parameters, namely the position, intensity and width of the main excitation, can be obtained as a function of the mean atomic number of the target $\left<Z_{\rm t}\right>$, that is, the number of electrons per formula divided by the number of atoms \cite{Tan2004}.
This approach has been shown to provide reasonable predictions of the optical ELF of arbitrary biological materials \cite{Tan2004,deVera2013,deVera2013b}.

The optical ELFs of the target materials have been extended to the whole energy and momentum plane (the Bethe surface), following Garcia-Molina, Abril \textit{et al.}, by means of the MELF-GOS (Mermin Energy-Loss Function--Generalized Oscillator Strengths) method \cite{Abril1998,Heredia-Avalos2005,Garcia-Molina2009a,Garcia-Molina2012b}. It uses Mermin functions to describe the outer-shell electrons and hydrogenic GOS to describe the inner-shells of atomic character. This method has demonstrated its ability to properly reproduce the experimental Bethe surface of liquid water \cite{Abril2010a} and to produce accurate electronic cross sections for a wide variety of target materials \cite{Heredia-Avalos2005,Garcia-Molina2009a,deVera2013,deVera2015}. It also fulfills a series of physical constraints unlike simpler dispersion algorithms \cite{Chantler2015}.
This methodology was applied previously for the calculation of electron MFPs in organic materials and metals~\cite{Denton_2008_SurfInterfaceAnal.40.1481, Garcia-Molina2016}.

\begin{figure}[t]
	\begin{centering}	
    \includegraphics[width=0.45\textwidth]{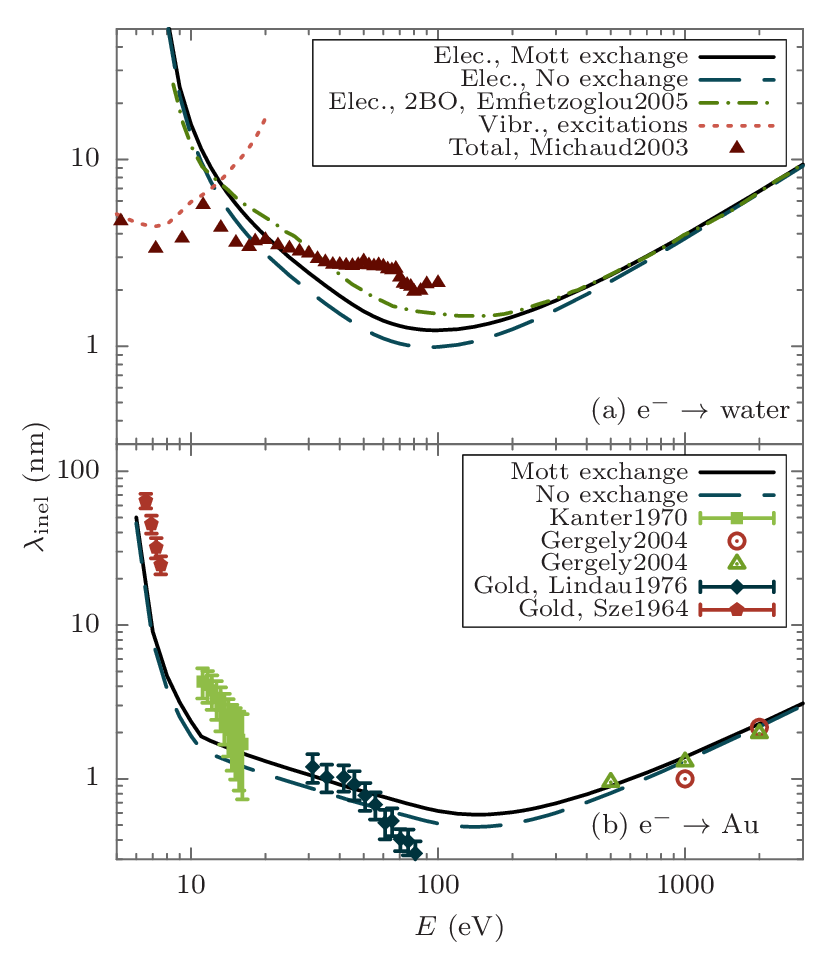}
	\par\end{centering}
	\caption{Inelastic mean free path $\lambda_\mathrm{inel}$ of electrons in (a) liquid water and (b) gold. Lines represent calculations within the dielectric formalism at different levels of approximation while symbols are experimental data for ice \cite{Michaud2003} and gold \cite{Sze1964,Lindau1976,Kanter1970,Gergely2004}.}
    \label{fig:inelMFPcalcs}
\end{figure}

The projectile electron is characterized by its kinetic energy $E$ and by the exchange factor $f_{\rm ex}$, accounting for the primary and secondary electron indistinguishability. In this work we have used the Mott exchange factor \cite{ICRU55}.

\begin{figure}[t]
 	\begin{centering}	
	\includegraphics[width=0.45\textwidth]{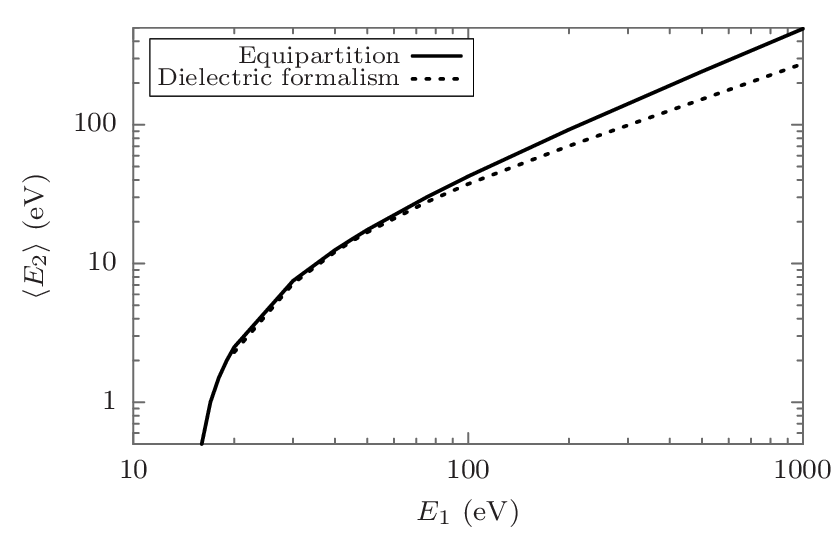}
	\par\end{centering}
	\caption{Average energy $\left< E_2 \right>$ of the electrons ejected by electron impact in liquid water versus projectile energy $E_1$. The dashed line represents the dielectric formalism calculations, while the solid line is the result of the equipartition approximation. See the text for further details.}
    \label{fig:W2}
\end{figure}

Calculations are benchmarked against experimental da\-ta and other models in Fig.~\ref{fig:inelMFPcalcs}. First let us analyze the effect of the exchange. The Mott factor increases the inelastic MFP, improving the agreement between the calculations for liquid water and the experiments for ice \cite{Michaud2003}. The same happens for the calculations for gold and the experimental data from \cite{Gergely2004}. At low energies, around 16~eV, the calculations corrected for exchange are still within the experimental uncertainties of the data from \cite{Kanter1970}.
A more sophisticated calculation for liquid water where second Born order corrections are included \cite{Emfietzoglou2005} is also shown. For electron energies around and below about 100~eV the second Born order corrections should, in principle, be included to use the dielectric formalism accurately \cite{Emfietzoglou2005}. However, the impact of this correction on the liquid water MFP seems to be relatively small. This fact has been observed for other materials \cite{Ashley1979,Tung1994,Tung2007}. Indeed, Lindhard \cite{Lindhard1954} and Ritchie \cite{Ritchie1975} pointed out that the first Born approximation might work well for calculating the MFP at any electron velocity. This is due to the fact that the relative velocity of an incident electron to a representative electron in the outer shells of the target is always large enough regardless of how small the incident electron velocity is and due to the screening of the projectile Coulomb field by the polarization of the target electron gas at low velocities.

Dielectric formalism calculations for both liquid water and gold begin to depart from experimental data at low energies, $E\leq 16$~eV. For gold this might be mainly related to the limitations of the dielectric model but for water, apart from these intrinsic limitations, other excitation channels become dominant at low energies. The dotted line in Fig. \ref{fig:inelMFPcalcs} shows the vibrational MFP calculated using the vibrational cross sections for the water molecule recommended by Itikawa and Mason \cite{Itikawa2005} on the base of a comprehensive compilation of experimental and theoretical data. It can be seen that the decrease of the experimental inelastic MFP for water \cite{Michaud2003} around 4~eV perfectly agrees with the minimum of the vibrational MFP, confirming that this inelastic channel becomes dominant at low energies. In this work we have summed the calculated electronic cross sections and the recommended vibrational cross sections in order to obtain an inelastic MFP which extends over a wide energy range. The result is the long-dashed line shown in Fig. \ref{fig:inelMFP}. For gold, due to the model uncertainties at low energies, we decided to make an interpolation of experimental data for $E\leq 16$~eV while using calculated data for larger energies.

Finally, the dielectric formalism can be extended to yield the energy spectrum and total ionization cross sections of biomaterials impacted by fast ions and electrons \cite{deVera2013,deVera2013b,deVeraThesis} by introducing a mean binding energy for the outer shell electrons $\left<B\right>$ so the energy of the ejected electrons is given by $E_2 = \Delta \varepsilon - \left<B\right>$. By using this methodology and using $\left<B\right> = 15$~eV and the Mott exchange we have calculated the average energy of electrons produced by electron impact in liquid water, as explained elsewhere \cite{deVera2013b,deVeraThesis}, in order to determine the energy of the electrons of the second generation.
In Fig.~\ref{fig:W2} the current results (dashed line) are compared to the approximation considered in Ref.~\cite{Surdutovich2015} (solid line) where the remaining energy after the electron ejection is equally distributed between the primary and the secondary electron.
It can be seen that this approximation is valid below 100~eV.
Below 15~eV it is not possible to produce the second generation of electrons, since these energies are below the assumed mean ionization energy. It should be noted that, although the present results are for liquid water, the calculation of $\left< E_2 \right>$ is not very sensitive to the composition of the biological target \cite{deVera2013b} and should therefore be valid for the present analysis.

\subsection{Partial wave analysis calculations for elastic cross sections\label{app:el_cross_sections}}

The scattering of electrons by atoms can be described by a central potential of the form \cite{Salvat2003,Salvat2005}:
\begin{equation}
V(r) = V_{\rm st}(r)+V_{\rm ex}(r)+V_{\rm cp}(r)-i W_{\rm abs}(r) \ ,
\label{eq:V}
\end{equation}
where $V_{\rm st}(r)$, $V_{\rm ex}(r)$, and $V_{\rm cp}(r)$ are the electrostatic, exchange, and correlation-polarization potentials, respective\-ly, and $W_{\rm abs}(r)$ is the magnitude of the imaginary absorption potential. The electrostatic potential represents the pure Coulomb interaction while $V_{\rm ex}(r)$ accounts for the electron indistinguishability and $V_{\rm cp}(r)$ accounts for electron correlation and induced polarization of the atom electron cloud. The absorption term accounts for the loss of electrons from the elastic channel due to inelastic collisions. This latter term is needed to reproduce experimental data in which the inelastic collisions are unavoidable.

The fact that the potential is spherical allows atom to be described using the direct scattering and the spin-flip scattering amplitudes
which depend on the polar scattering angle $\theta$ as determined from the large-$r$ behavior of the Dirac distorted plane waves, i.e. the solutions of the Dirac equation for the central potential $V(r)$ that behave asymptotically as a plane wave \cite{Salvat2003,Salvat2005}. The scattering amplitudes can be found by partial-wave expansion and then the differential elastic cross section (DCS) can be calculated as well the total elastic cross section (and the inverse mean free path) by integrating the DCS over the scattering angle.

We have used the code ELSEPA \cite{Salvat2005} to obtain the elastic cross sections. ELSEPA allows cross sections to be calculated down to incident electron energies of 10~eV. In general the electrostatic and exchange terms are the dominant components of the potential, although the correlation-pola\-rization term becomes very important at low electron energies, as shown below. For the present calculations we have used, for each interaction, the potentials recommend\-ed by \cite{Salvat2005}. The electrostatic contribution is calculated by accounting for the nucleus charge distribution by a Fermi distribution, while the electronic structure is obtained from numerical Dirac–-Fock distribution read from the database file. The Furness-McCarthy potential is used for the exchange contribution. For the correlation-polariza\-tion the local density approximation (LDA) is used for electron correlation together with an asymptotic potential for polarization by distant collisions. The absorption term can be also taken into account within the LDA \cite{Salvat2005}.

To test the code we have performed calculations of differential elastic cross section (DCS, angular distribution) for low-energy electrons scattered by water molecules and Hg atoms. The latter case is the situation we found in the literature closest to our present problem of LEEs colliding with heavy atoms, such as gold, where experimental data is available \cite{Holtkamp1987,Panajotovic1993}. Experimental information is also available for water \cite{Cho2004}.
We find that elastic cross sections are strongly affected by the inclusion of the polarization potential.
We also included the absorption potential for these tests, since it is needed to reproduce the experimental results, where the loss of electrons from the elastic channel is unavoidable. The general trends of the DCS are in fair agreement with the experimental data.

\begin{figure}[t]
	\begin{centering}	
    \includegraphics[width=0.45\textwidth]{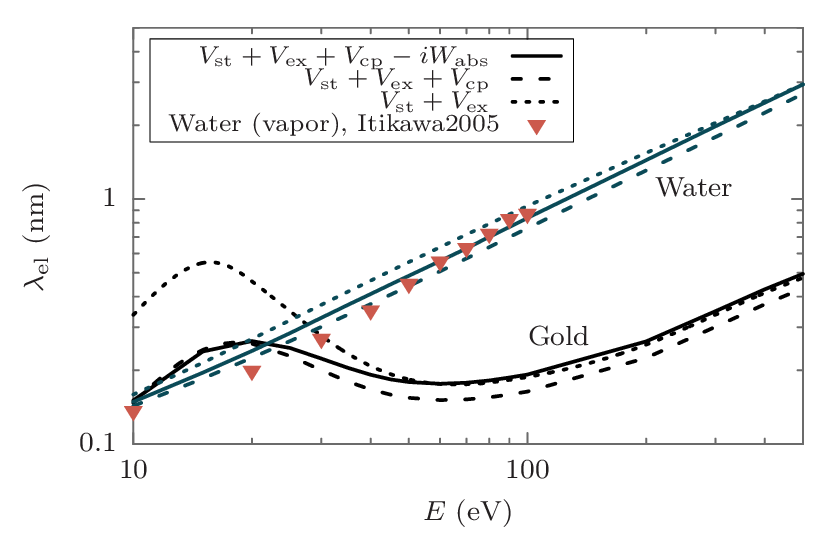}
	\par\end{centering}
	\caption{Elastic mean free path $\lambda_\mathrm{el}$ of electrons in liquid water and gold. Lines represent partial wave analysis calculations with several contributions from Eq.~\eqref{eq:V} while symbols are recommended data for water (scaled to liquid water density) \cite{Itikawa2005}.
    }
    \label{fig:elMFPcalcs}
\end{figure}

Figure \ref{fig:elMFPcalcs} shows the calculated elastic MFP for electrons in liquid water and gold. Dashed lines represent calculations where only the electrostatic and exchange potentials have been included (basic ELSEPA calculation), while solid lines also include the correlation-polarization potential and dotted lines include the absorption contribution. Symbols are the recommended values for water given by \cite{Itikawa2005}, based on an extensive compilation of experimental and theoretical data. The values for water vapor have been scaled to liquid water density. As can be seen, the correlation-polarization potential has a great impact at low energies, especially for gold. If not included the gold MFP is much lower than for water at high energies, while it becomes larger at low energies. However, the inclusion of the correlation-polarization potential makes the gold MFP converge with that of water at low energies.
The results for water are in fairly good agreement with the recommended values from Ref. \cite{Itikawa2005}. As expected the inclusion of the absorption potential slightly increases the agreement, since loss of electrons from the elastic channel by inelastic interactions is impossible to avoid in experiments. Both for gold and for water the absorption slightly increases the absolute values of the MFP, although the general shape is not modified. We stress that, for feeding the diffusion equation, we do not include the absorption term in the elastic MFP calculation but inelastic interactions are already included in the inelastic MFP, so only elastic interactions should be included in the elastic MFP.

\bibliographystyle{epj}
\bibliography{library}

\end{document}